\documentclass{aa}

\usepackage[varg]{txfonts}

\bibpunct{(}{)}{;}{a}{}{,} 

\usepackage{graphicx}

\usepackage{textcomp}

\usepackage{multirow}

\begin{document}

        \title{Comparison of space weathering spectral changes induced by solar wind and micrometeoroid impacts using ion- and femtosecond-laser-irradiated olivine and pyroxene\thanks{All measured spectra in raw format are only available in electronic form at the CDS via anonymous ftp to cdsarc.u-strasbg.fr (130.79.128.5) or via http://cdsweb.u-strasbg.fr/cgi-bin/qcat?J/A+A/}}
        
        \author{Kate\v rina Chrbolková\inst{\ref{inst1},\ref{inst2},\ref{inst8}}
        \and Rosario Brunetto\inst{\ref{inst3}}
        \and Josef \v Durech\inst{\ref{inst1}}
        \and Tomá\v s Kohout\inst{\ref{inst2},\ref{inst8}}
        \and Kenichiro Mizohata\inst{\ref{inst4}}
        \and Petr Malý\inst{\ref{inst6}}
        \and Václav D\v edi\v c\inst{\ref{inst7}}
        \and Cateline Lantz\inst{\ref{inst3}}
        \and Antti Penttil\" a\inst{\ref{inst5}}
        \and Franti\v sek Trojánek\inst{\ref{inst6}}
        \and Alessandro Maturilli\inst{\ref{inst9}}
        }
        
        \institute{Astronomical Institute, Faculty of Mathematics and Physics, Charles University, V Hole\v sovi\v ckách 2, 18000, Prague, Czech Republic \email{katerina.chrbolkova@helsinki.fi}\label{inst1}
        \and 
        Department of Geosciences and Geography, PO box 64, 00014 University of Helsinki, Finland\label{inst2} 
        \and
        Czech Academy of Sciences, Institute of Geology, Rozvojová 269, 16500, Prague, Czech Republic\label{inst8}
        \and 
        Université Paris-Saclay, CNRS, Institut d'Astrophysique Spatiale, 91405 Orsay, France\label{inst3}
        \and
        Department of Physics, PO box 43, 00014 University of Helsinki, Finland\label{inst4}
        \and 
        Department of Chemical Physics and Optics, Faculty of Mathematics and Physics, Charles University, Ke Karlovu 3, 12116 Prague, Czech Republic\label{inst6}
        \and 
        Institute of Physics, Faculty of Mathematics and Physics, Charles University, Ke Karlovu 5, 12116 Prague, Czech Republic\label{inst7}
        \and
        Department of Physics, PO box 64, 00014 University of Helsinki, Finland\label{inst5}
        \and
        Institute of Planetary Research, DLR German Aerospace Centre, Rutherfordstrasse 2, 12489 Berlin, Germany\label{inst9}
        }

        \abstract 
        {
                Space weathering is a process that changes the surface of airless planetary bodies. Prime space weathering agents are solar wind irradiation and micrometeoroid bombardment. These processes alter planetary reflectance spectra and often modify their compositional diagnostic features.
        } 
        {
                In this work we focused on simulating and comparing the spectral changes caused by solar wind irradiation and by micrometeoroid bombardment to gain a better understanding of these individual space weathering processes.
        } 
        {
                We used olivine and pyroxene pellets as proxies for planetary materials. To simulate solar wind irradiation we used hydrogen, helium, and argon ions with energies from 5 to 40 keV and fluences of up to $10^{18}$ particles/cm$^2$. To simulate micrometeoroid bombardment we used individual femtosecond laser pulses. We analysed the corresponding evolution of different spectral parameters, which we determined by applying the Modified Gaussian Model, and we also conducted principal component analysis.
        } 
        {
                The original mineralogy of the surface influences the spectral evolution more than the weathering agent, as seen from the diverse evolution of the spectral slope of olivine and pyroxene upon irradiation. The spectral slope changes seen in olivine are consistent with observations of A-type asteroids, while the moderate to no slope changes observed in pyroxene are consistent with asteroid (4) Vesta. We also observed some differences in the spectral effects induced by the two weathering agents. Ions simulating solar wind have a smaller influence on longer wavelengths of the spectra than laser irradiation simulating micrometeoroid impacts. This is most likely due to the different penetration depths of ions and laser pulses. Our results suggest that in some instances it might be possible to  distinguish between the contributions of the two agents on a weathered surface.
        }
        {}
        
        \keywords{planets and satellites: surfaces - (Sun:) solar wind - meteorites, meteors, meteoroids - methods: data analysis - techniques: spectroscopic}
        
        \titlerunning{Comparison of space weathering spectral changes}
        
\maketitle

\section{Introduction}
The surfaces of airless planetary bodies in the solar system are permanently exposed to the space environment, which consists  mainly of solar wind irradiation and micrometeoroid bombardment. These processes cause changes to the topmost surface layers of the bodies, resulting in an alteration of their spectroscopic features \citep{hapke_65,hapke_01,wehner_63}. The exact understanding of the individual contribution of these two processes to the alteration of reflectance spectra is of utmost importance, and is the focus of our work.

Among the most prominent visible (VIS) and near-infrared (NIR) spectral changes observed on dry silicate bodies, such as the Moon or S-type asteroids, are spectral slope reddening (a decrease in reflectance towards shorter wavelengths), reduction of the diagnostic 1 and 2 \textmu m absorption bands \citep{burns_89}, and an overall darkening of spectra  \citep[see e.g.][]{hapke_01,pieters_00,pieters_16}.

Several silicate diagnostic features are also located in the mid-infrared (MIR) wavelength range, such as the Christiansen feature (around 8.5 \textmu m) or the fundamental molecular vibration bands, the reststrahlen bands \citep{salisbury_91}. Space weathering does not influence these in the same manner as those at VIS and NIR wavelengths. The most prominent MIR changes are the alteration of the shape, position, and relative intensities of the reststrahlen bands \cite[see e.g.][]{lucey_17,brunetto_20}. 

The spectra of minerals are influenced by many effects related to  space weathering, such as melting, vaporization and associated vapour redeposition, sputtering, particle size segregation, blistering, and amorphization. Many of the spectral changes are attributed to the creation of  nanophase iron particles \citep{hapke_01,noguchi_14,matsumoto_15,pieters_00}. These particles are created when an iron (Fe) ion is freed from the silicate crystalline structure during solar wind or micrometeoroid bombardment. 

To evaluate and understand the spectral changes induced by space weathering, previous studies compared asteroid spectra with those of fresh compositionally identical meteorites. Additionally, many studies subjected meteorite and planetary analogue materials to laboratory simulations that mimic space weathering processes. Widely used techniques include ion irradiation, which simulates solar wind, and laser irradiation, which simulates micrometeoroid impacts.

Ion irradiation experiments have mostly verified the general trends of spectral darkening and reddening at VIS and NIR wavelengths. Several works related to silicate materials have given insights into space weathering, including  \cite{brunetto_06}. By irradiating olivine and pyroxene samples and comparing their spectra with the spectrum of asteroid (832)~Karin, they estimated its surface exposure age, which is in accordance with Karin's collisional history. Another example is the work of \cite{fulvio_12}, who simulated weathering on asteroid (4) Vesta by irradiation of eucrite meteorites. They found that different parts of the meteorite weather slightly differently. Based on their work, the diversity of V-type asteroids can be explained by distinct space weathering stages (for more examples of ion irradiation experimental work, see \citealt{lantz_14, lantz_17, loeffler_09, paillet_03, vernazza_13}).

Among the most prominent results based on laser experiments is the observation that the degree of weathering depends on the starting mineralogy of the processed sample \citep{yamada_99}. Olivine shows much greater changes than pyroxene, and these changes are not linear with respect to time \citep{sasaki_02}. Laser-irradiated ordinary chondrites can reproduce spectra of real asteroids, as shown by \cite{sasaki_02}, among others, for asteroids (349) Dembowska and (446) Aeternitas. After comparing laser-irradiated pyroxene samples with asteroid (4) Vesta, \cite{yamada_99} concluded that Vesta is more weathered than their samples (for additional examples, see \citealt{gillis-davis_18, matsuoka_20, moroz_96, jiang_19, sasaki_01b}).

These laser irradiation experiments were mostly based on the irradiation effects induced by employing nanosecond pulsed lasers. The main argumentation in favour of this setting is that a duration of several nanoseconds approximately matches the duration of micrometeoroid impacts \citep{sasaki_02, yamada_99}. In addition, the ion yield created by irradiation is similar to the real situation in the case of high enough irradiation fluences \citep{kissel_87}. Recent work of \cite{fulvio_21} shows that morphology of the craters formed by nanosecond pulses resembles that induced by natural micrometeorite impacts. On the other hand, this set-up causes mainly heating and melting with subsonic and sonic evaporation from the surface \citep{gusarov_05}. In contrast, a femtosecond laser pulse with high peak irradiance allows the propagation of a shock wave of several tens of GPa and confined melting \citep{berthe_11, boustie_08}. During such short pulses the laser beam does not interact with the created vapour plume. As a consequence of this set-up, ablation of the material proceeds through spallation, fragmentation, homogeneous nucleation, and vaporization \citep{perez_03}, resulting in a subsurface structure of the craters that is remarkably similar to that of the Moon and of asteroid (25143)~Itokawa \citep{fazio_18}; the results are in accordance with the findings of \cite{fiege_19} regarding the  destruction of   material by micrometeoroid impacts. \cite{shirk_99} also pointed out that the ablation depth is larger in the case of femtosecond (compared to nanosecond) regime due to increased absorption coefficient.

Despite these experimental efforts, the difference between the spectral effects of the solar wind and micrometeorite impacts is not well understood. Most of the laboratory work conducted to date has only  focused  on one specific space weathering process, either ion or laser bombardment. One of the exceptions is the work of \cite{gillis-davis_18}, who showed that laser and electron irradiation of the Murchison meteorite separately yield  small changes, but when combined the changes become significant. Another example is the comparison of ion and laser irradiation of olivine done by \cite{loeffler_09}. They found that different weathering mechanisms can lead to similar effects on the spectral curves of olivine-rich asteroids.

To analyse the differences induced by solar wind and micrometeoroid bombardment on planetary surfaces, we performed ion and laser irradiation experiments on the typical silicate minerals olivine and pyroxene. These minerals can be found not only in the S- and Q-type asteroids but also in V- and A-types and across the lunar surface (see \citealt{binzel_04, demeo_09, deleon_06, desanctis_11, gaffey_93, keller_14, sanchez_14, sunshine_07, burns_72} and \citealt{isaacson_10} for the occurrence of these minerals on asteroid bodies and the Moon). 

\section{Methods}
We performed two different types of irradiation experiments on terrestrial silicates also found on planetary surfaces: ion and femtosecond laser irradiation. For ion irradiation we used three different types of ions: hydrogen (H$^+$), helium (He$^+$), and argon (Ar$^+$). Laser irradiation was done using individual femtosecond laser pulses. Our experiments were conducted at three different institutes, but all of them were done using the same sample material (see Sect. \ref{methods-samples_sect}), which gives a great opportunity for comparison of spectral signatures under different irradiation conditions. The experimental conditions are described in detail in Sects. \ref{methods-ion_sect} and \ref{methods-laser_sect}. For an overview of the experimental set-up, see Table~\ref{experimental_setup_table}. For additional notes on experimental set-up, see Appendix~\ref{appendix_setup}. 
\begin{table*}
        \caption{Overview of the experimental set-up.}
        \label{experimental_setup_table} 
        \centering
        \begin{tabular}{c c c c c c c c}   
                \hline\hline
                \multicolumn{2}{c}{Irradiation type} & Irradiation (ions/cm$^2$) & $p$ (mbar) & Spectroscopy & Spectrometer set-up & Place \\
                \hline 
                \multirow{2}{*}{H$^+$}  &  \multirow{2}{*}{OL} & $10^{14}$, $10^{15}$, $10^{16}$, $10^{17}$, &  \multirow{2}{*}{$10^{-7}$} &  \multirow{2}{*}{ex situ} &  \multirow{2}{*}{Integrating sphere, $i=10^{\circ}$} &  \multirow{2}{*}{HEL} \\
                & &  $2\times 10^{17}$, $5\times 10^{17}$, $10^{18}$ & & & &\\
                \rule{0pt}{3ex}H$^+$    & OPX & $10^{16}$, $10^{17}$, $2\times 10^{17}$, $5\times 10^{17}$, $10^{18}$ & $10^{-7}$ & ex situ & Integrating sphere, $i=10^{\circ}$ & HEL \\
                \rule{0pt}{3ex} \multirow{2}{*}{He$^+$} & \multirow{2}{*}{OL, OPX} & \multirow{2}{*}{$10^{16}$, $3\times10^{16}$, $6\times10^{16}$, $10^{17}$} & \multirow{2}{*}{$10^{-7}$} & \multirow{2}{*}{in situ} &   $i_{\textrm{V}}=c_{\textrm{V}}=15^{\circ}$, $\phi_{\textrm{V}}=20^{\circ}$ & \multirow{2}{*}{PAR} \\
                & & & & & $i_{\textrm{N}}=20^{\circ}$, $c_{\textrm{N}}=\phi_{\textrm{N}}=15^{\circ}$ & \\
                \rule{0pt}{3ex} \multirow{2}{*}{Ar$^+$} & \multirow{2}{*}{OL, OPX} & $10^{15}$, $3\times10^{15}$, $6\times10^{15}$, $10^{16}$, & \multirow{2}{*}{$10^{-7}$} & \multirow{2}{*}{in situ} &  $i_{\textrm{V}}=c_{\textrm{V}}=15^{\circ}$, $\phi_{\textrm{V}}=20^{\circ}$  & \multirow{2}{*}{PAR} \\
                & &  $2\times10^{16}$, $6\times10^{16}$, $10^{17}$ &  &  &   $i_{\textrm{N}}=20^{\circ}$, $c_{\textrm{N}}=\phi_{\textrm{N}}=15^{\circ}$  & \\
                \hline 
                & & Irradiation (J/cm$^2$) & & &  & \\
                \hline
                \multirow{2}{*}{Laser} & \multirow{2}{*}{OL} & 1.7, 2.4, 3.8, 4.6, 6.7, 10.4,  & \multirow{2}{*}{$10^{-4}$} & \multirow{2}{*}{ex situ} & \multirow{2}{*}{$i =0^{\circ}$, $c=30^{\circ}$} & \multirow{2}{*}{PRG} \\
                &       & 15, 23.4, 30.6, 60, 93.8, 375 &  &  & & \\
                \rule{0pt}{3ex} \multirow{2}{*}{Laser} &  \multirow{2}{*}{OPX}  & 4.5, 5.6, 12.5, 18, 28.1, 36.7, &  \multirow{2}{*}{$10^{-4}$} &  \multirow{2}{*}{ex situ} &  \multirow{2}{*}{$i =0^{\circ}$, $c=30^{\circ}$} &  \multirow{2}{*}{PRG} \\
                & & 50, 72, 112.5, 200, 450, 1800 & & & & \\
                \hline 
        \end{tabular}
        \tablefoot{OL and OPX stand for olivine and pyroxene, respectively; H$^+$, He$^+$, and Ar$^+$ correspond to hydrogen, helium, and argon ions, respectively; $p$ is the pressure in the vacuum chamber during the irradiation; $i$ is the incidence, $c$ the collection, and $\phi$ the phase angle of the spectral measurements; the subscripts stand for different wavelength ranges: V is the visible and N is the near-infrared. HEL indicates the University of Helsinki laboratories, PAR indicates the INGMAR set-up at IAS-CSNSM in Orsay, and PRG indicates the laboratories at Charles University in Prague.}
\end{table*}

\subsection{Samples}
\label{methods-samples_sect}
We acquired a bulk sample of aluminium-rich enstatite (pyroxene), including $\approx 7\,\%$ pargasite, collected in M\"antyharju, Finland. Samples obtained from it are denoted  OPX (OrthoPyroXene). We also used olivine pebbles from Hebei Province, China, as in \cite{yazhou_17}. Samples from the olivine pebbles are denoted  OL. Based on electron probe microanalysis using CAMECA SX-100, the enstatite content\footnote{Calculated as a ratio of Fe to magnesium (Mg) content: Fe/(Fe+Mg)*100.} of the OPX sample was $67.1\pm1.6$ and the forsterite content of the OL sample was $90.1\pm0.8$, evaluated from seven and nine measurements, respectively. These values are similar to H chondrite meteorites and diogenites \citep{cambr_02}. For a discussion of possible contamination of our samples and its influence on the final results, see Appendix \ref{appendix_contamination}.

Each of the materials was ground using a mortar and pestle, and consequently dry-sieved to $<\,106\,$\textmu m, which can be considered a suitable analogue of the regolith of planetary surfaces. The mean size of lunar regolith is around 60 \textmu m \citep{mckay_78,pieters_00}, and its optical properties are, according to \cite{pieters_00}, dominated by regolith grains $<\,45\,$\textmu m. Our choice of sizes thus covers the most relevant size fraction for spectroscopic studies. Additionally, the sample also includes particles larger than the already mentioned $45\,$\textmu m to give us a more relevant picture of a planetary surface. Particles larger than 100 \textmu m are more than 100 times larger than the average wavelength of our VIS-NIR measurements. This means that they are already outside the so-called resonance size range regarding their light scattering behaviour and are within the geometric optics regime. At these sizes the light reacts with particle surfaces and subsurface volumes rather than with the particle as a whole, and the changes in particle size does not influence the distribution of the reflected light much. Including larger particles in our samples is thus not needed.

Using the Specac manual hydraulic press, we prepared two  sets of pellets (OL and OPX) with a diameter of 13 mm. Each pellet was made of 100~mg of the mineral on top of 700~mg of potassium bromide (KBr) substrate. Using a KBr substrate ensures greater durability of the pellet. Still, the mineral layer on top is optically thick enough to enable analysis of only the mineral material. The approximate thickness of the mineral layer was 200~\textmu m. The pellets were created using a pressure of 6~t for 6~min ($\approx 450$~MPa), which guarantees compactness of the material, while the mineral surface stays in a loose form that can be blown away or wiped off with a finger. This approach has already been used in the work of \cite{brunetto_14} and \cite{lantz_17}, who also discuss the suitability of   using the  pellets for solar system studies. Some of the pellets are shown in Fig.~\ref{evolution_fig}.

\subsection{Ion irradiation}
\label{methods-ion_sect}
He$^+$ and Ar$^+$ irradiation was conducted using the INGMAR set-up (IAS-CSNSM, Orsay) interfaced to the SIDONIE implanter (CSNSM, Orsay; see \citealt{chauvin_04, brunetto_14}). The pellets were placed into a vacuum chamber (pressure $\approx\,10^{-7}$ mbar) for the whole experiment, that is the irradiation and spectral measurements. He$^+$ irradiation was done with 20 keV ions, simulating ions from solar active regions and solar energetic particles in the low-energy range. The energy of the ions is approximately five times greater than that of the standard solar wind. Nevertheless, by using argumentation similar to that of \cite{brunetto_14}, we can conclude that timescales corresponding to our irradiation fluences are correct within an order of magnitude to the standard solar wind situation. We used four fluences: $10^{16}$, $3\times10^{16}$, $6\times10^{16}$, and $10^{17}$~ions/cm$^2$. In the case of Ar$^+$, we used 40~keV ions to simulate the effects of heavy ions in the slow solar wind, using seven different fluences: $10^{15}$, $3\times10^{15}$, $6\times10^{15}$, $10^{16}$, $2\times10^{16}$, $6\times10^{16}$, and $10^{17}$~ions/cm$^2$. The ion beam was scanned vertically and horizontally across the surface to produce homogeneous irradiation \citep{chauvin_04}. Spectral measurements were performed in situ in the VIS wavelength range using a Maya2000\,Pro (OceanOptics) and in the NIR range with a Tensor\,37 (Bruker). The illumination angle for VIS measurements was 15$^\circ$ and the collection angle 15$^\circ$. The phase angle was, due to the three-dimensional geometry, 20$^\circ$. In the NIR wavelength range the illumination angle was 20$^\circ$, the collection angle was the same as for VIS measurements, and the phase angle thus equalled 15$^\circ$. The collection spot for VIS-NIR measurements was $\approx\,4$~mm on the surface of the sample pellet and the Spectralon, which was used as a calibration standard.

To simulate irradiation by the slow solar wind, we conducted the H$^+$ irradiation at the Helsinki Accelerator Laboratory, Finland, where we used 5~keV protons. As in the above-mentioned case of He$^+$ irradiation, our results can be applied to the standard solar wind situation. We irradiated using seven different fluences: $10^{14}$, $10^{15}$, $10^{16}$, $10^{17}$, $2\times 10^{17}$, $5\times 10^{17}$, and $10^{18}$~ions/cm$^2$ in the case of OL samples and five different fluences: $10^{16}$, $10^{17}$, $2\times 10^{17}$, $5\times 10^{17}$, and $10^{18}$~ions/cm$^2$ in the case of OPX samples. During irradiation the samples were placed in a vacuum chamber ($\approx 10^{-7}$~mbar), but the spectral measurements were performed ex situ. As was noted, for example, by \cite{loeffler_09}, after the sample is taken out of the vacuum chamber in which the irradiation was taking place, reoxidation of the created metallic Fe nanoparticles takes place within minutes. This may erase part of the spectral changes. Before extraction from the vacuum chamber, we thus filled the chamber with a mixture of nitrogen and 2\% oxygen gas, which guaranteed that only a thin surface layer of the sample oxidized protecting the underlying nanophase Fe particles (for more details on surface passivation, see \citealt[][and their scanning transmission electron microscopy analysis of nanophase Fe particles)]{kohout_14}. The hemispherical reflectance spectra were measured in the shortest possible time after the irradiation using an OL-750 automated spectroradiometric measurement system by Gooch \& Housego located at the Department of Physics, University of Helsinki \citep{penttila_18}. The OL-750 instrument is equipped with polytetrafluoroethylene (PTFE) and golden integrating spheres and with a specular reflection trap. The spectra were measured relative to PTFE (VIS) and golden (NIR) standards. The spectral resolution of the instrument varied between 5 and 10~nm. The spectrometer had an identical viewing geometry for all the segments of the spectra and an incidence angle of 10$^{\circ}$ to the surface normal. The size of the collection spot on the sample surface was $\approx\,6$~mm for both the segments.

\subsection{Femtosecond laser irradiation}
\label{methods-laser_sect}
We conducted the femtosecond laser irradiation at the Department of Chemical Physics and Optics, Charles University, the Czech Republic. Our set-up was inspired by the work of \cite{fazio_18}. We used a titanium sapphire Tsunami laser and Spitfire amplifier, Spectra-Physics, Newport. The laser wavelength was 800~nm and it shot individual 1.5~mJ (for OL) and 1.8~mJ (for OPX) 100~fs pulses into a square grid ($3\times3$ mm). The spot size on the surface of the pellet was $\approx\,50$~\textmu m, as estimated by the razor blade method (see e.g. \citealt{khosrofian_83}). The final energy surface density (further noted as energy density) of one pulse was then $\approx$\,80~J/cm$^2$. Considering a dust particle with a diameter of 1~\textmu m and typical velocity of 10~km/s, the energy density of our experiments is comparable to that of a micrometeoroid impact ($\approx$\,32~J/cm$^2$). By varying the spatial distance between the adjacent pulses we simulated different levels of space weathering;  the higher the density of the pulses, the higher the simulated space weathering level. The lowest coverage corresponds to one shot every 300~\textmu m, that is 100 pulses in our 3 mm square, and the highest  to one shot every 10~\textmu m, resulting in 9\,000 pulses in a square. In the case of OL we used the following irradiations (recalculated to the energy density, calculated as a number of pulses in cm$^2$ times energy of one pulse): 1.7, 2.4, 3.8, 4.6, 6.7, 10.4, 15.0, 23.4, 30.6, 60.0, 93.8, and 375.0 J/cm$^2$, and in the case of OPX: 4.5, 5.6, 12.5, 18.0, 28.1, 36.7, 50.0, 72.0, 112.5, 200.0, 450.0, and 1800.0 J/cm$^2$. We accommodated four such squares onto a pellet, with one of them always left unirradiated as a reference of the fresh state. The irradiation was conducted in a vacuum chamber ($\approx\,10^{-4}$~mbar). Samples then had to be taken out of the chamber as the spectral measurements were done in a different laboratory, for which  we surface passivated them as in the case of the H$^+$ irradiation (see Sect. \ref{methods-ion_sect}). 

The VIS and NIR spectral measurements were made using a Vertex 80v, Bruker, with an A513/Q variable angle reflection accessory using the Spectralon standard. The incidence angle was set to 0$^\circ$ and the collection angle to 30$^\circ$ to comply with Reflectance Experiment Laboratory (RELAB) database spectra. The vacuum in the spectrometer chamber was of the order of several mbar. 

With this spectrometer, we measured not only the VIS and NIR spectra but also the MIR spectra of all the samples (even those irradiated by ions) up to 13~\textmu m. The collection spot on the sample's surface was $\approx\,1.5$~mm for all the segments of the spectra, which ensured that only the irradiated part of the sample ($3\times3$ mm square) was measured but that it still contained at least (in the case of the least dense irradiation) 16 laser-induced craters. We did not have the commonly used calibration standard for these measurements; we thus used a silver-coated diffuse reflector from Thorlabs, which has a relatively uniform and high reflectance ( > 95 \%) in the MIR wavelength range. Consequently, we measured the diffuse reflector against the golden standard using the same observation geometry and the same set-up at the German Aerospace Centre to get the absolute calibration. The absolute values may thus in the case of MIR measurements be slightly offset; we therefore use only relative values to make conclusions in this case. 

All segments of all the measured spectra in absolute values are available at the Centre de Données astronomiques de Strasbourg (CDS).

\subsection{Spectral fits}
\label{fitting_sect}
To obtain spectral parameters from the measured VIS--NIR spectra, we used the Modified Gaussian Model (MGM) by \cite{sunshine_90,sunshine_99}. The MGM takes a spectrum and an input file with user-estimated spectral parameters, and  iteratively searches for the best fit of the spectrum, in our case realized by a second-order polynomial continuum and a set of absorption bands, represented by modified Gaussian curves, all in the natural logarithm of reflectance. The choice of second-order polynomial is supported, for example, by the research of \cite{clenet_11} or \cite{han_20}. The depth of individual absorption bands is expressed by the band strength parameter, which corresponds to the amplitude of the modified Gaussian curve.

Pyroxene exhibits two characteristic absorption bands at around 1 and 2 \textmu m \citep{burns_89}. Pargasite inclusions show a weak absorption at 0.9 \textmu m, and another set of absorptions mainly between 2.2 and 2.4 \textmu m \citep{bhatt_17,rommel_17}. These are minor compared with the strong OPX absorptions, but we still fitted them, so the root-mean-square error decreases (see arrow in Fig.~\ref{spectra_fig} for the only visible pargasite band). 

In OL we observe three deep absorptions around 1~\textmu m \citep{burns_70}. These absorptions are so close to each other that they overlap in the resulting spectrum. 

\subsection{Principal component analysis}
\label{pca_methods}
To compare the results of ion- and laser-irradiated spectra, we also applied principal component analysis (PCA) to them. This dimensionality reduction technique takes an $m$-dimensional cloud of points, where $m$ is the number of wavelengths in the spectral data, and transforms it so that in the new orthogonal coordinate space the first coordinate (first principal component, PC1) corresponds to the direction in which the original cloud showed the greatest variance, the second coordinate (PC2) represents the direction with the second greatest variance, and so on. The new coordinate system usually does not have an exact physical meaning, but its usefulness has been proven in previous research: for example, for asteroid taxonomy based on PCA, see \cite{demeo_09}; for analysis of meteorite spectra, see \cite{penttila_18}; for research on the lunar magnetic anomalies, see \cite{chrbolkova_19} and \cite{kramer_11}; and  for PCA applied to the spectra of galaxies, see \cite{connolly_14}.

We applied PCA to our measured spectra to obtain more information on spectral differences. Furthermore, \cite{demeo_09} published an article based on a  dataset of the spectra of 371 asteroids measured from 450 to 2500~nm. These asteroids were assigned spectral types. For our analysis (based on silicate minerals) we extracted S-, Q-, V-, and A-type asteroids from this dataset. The principal component space defined by the measured irradiated pellets can be used as a basis for the asteroid spectra. The only constraint is that the asteroid spectra have the same format (wavelength sampling and range) as the original measured spectra. We thus obtained the principal component transformation of the asteroid spectral curves and could compare them with the irradiation evolutions.

\section{Results}
\label{results_section}
\begin{figure*}
        \resizebox{\hsize}{!}{\includegraphics{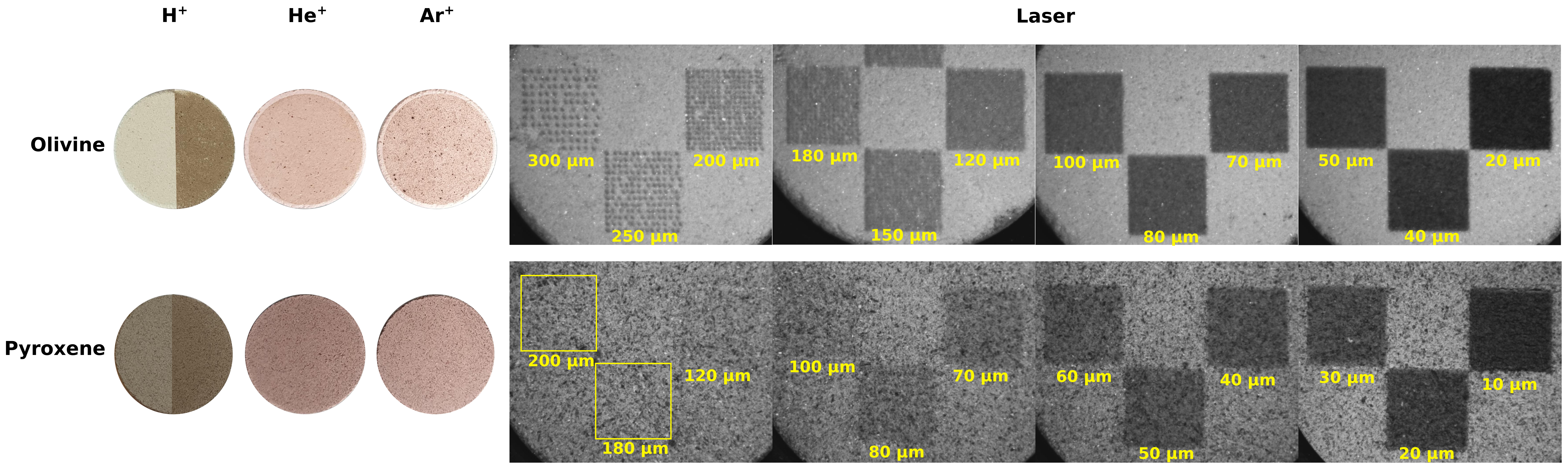}}
        \caption{Evolution of olivine and pyroxene pellets due to ion and laser irradiation. Pellets irradiated with hydrogen (H$^+$)  are photocomposites of the fresh surface in the left part and the $10^{18}$ ions/cm$^2$ irradiation in the right part. Pellets irradiated with helium (He$^+$) and argon (Ar$^+$) display the surface after $10^{17}$ ions/cm$^2$ irradiation; only the outermost annulus represents the fresh surface and was created by using a mask during irradiation. Yellow squares ($3\times3$ mm in size) in the photograph of a pyroxene pellet irradiated by laser indicate the regions of the least dense irradiation that are not visible by eye. The yellow numbers in the figure represent the spacing between two consecutive laser shots; for equivalent values of energy densities, see Table~\ref{experimental_setup_table}: the sequence of the distances relevant to one material in this figure (from left to right) corresponds exactly to the list in the table.}
        \label{evolution_fig}
\end{figure*}

\begin{figure*}
        \resizebox{\hsize}{!}{\includegraphics{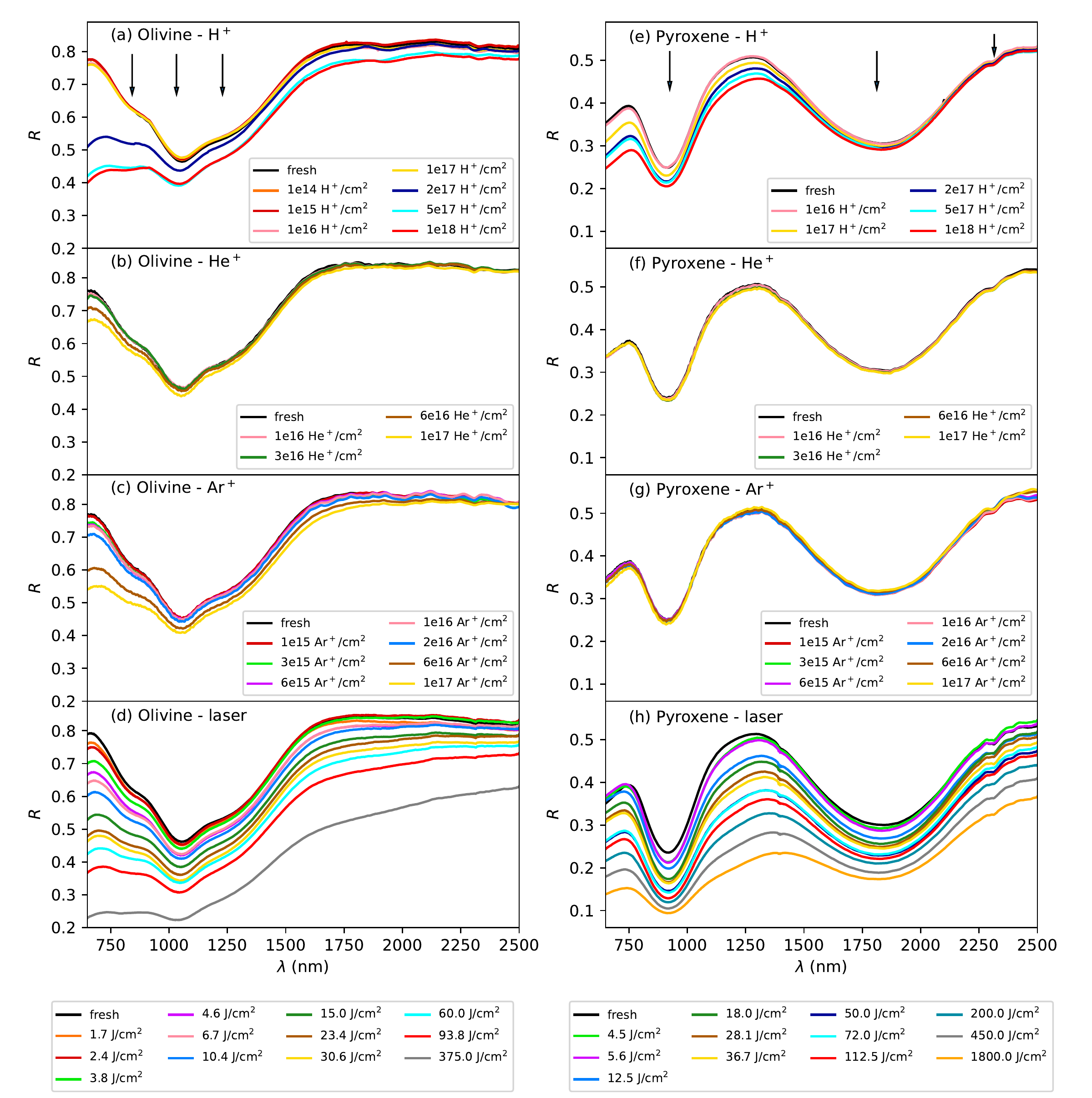}}
        \caption{Evolution of spectra due to different irradiation methods. Left: Olivine irradiated by (a) hydrogen (H$^+$), (b) helium (He$^+$), and (c) argon (Ar$^+$) ions, and (d) femtosecond pulsed laser with various ion fluences (ions/cm$^2$) or energy densities (J/cm$^2$) (see Sect. \ref{methods-laser_sect} for more details). Right: (e--h) pyroxene undergoing the same processing described for olivine in the left panel. In the graphs $R$ stands for reflectance and $\lambda$ is the wavelength. The arrows in (a) and (e) indicate the position of the diagnostic mineral bands; the smallest arrow in (e) refers to the position of the only visible pargasite band.}
        \label{spectra_fig}
\end{figure*}

We conducted the irradiation as described in Sects. \ref{methods-ion_sect} and \ref{methods-laser_sect}. Examples of the visual changes of the pellets' surfaces are shown in Fig.~\ref{evolution_fig}. The spectra resulting from the irradiation are displayed in Fig.~\ref{spectra_fig}.

\subsection{Time and energy evolution}
\begin{figure}
        \resizebox{\hsize}{!}{\includegraphics{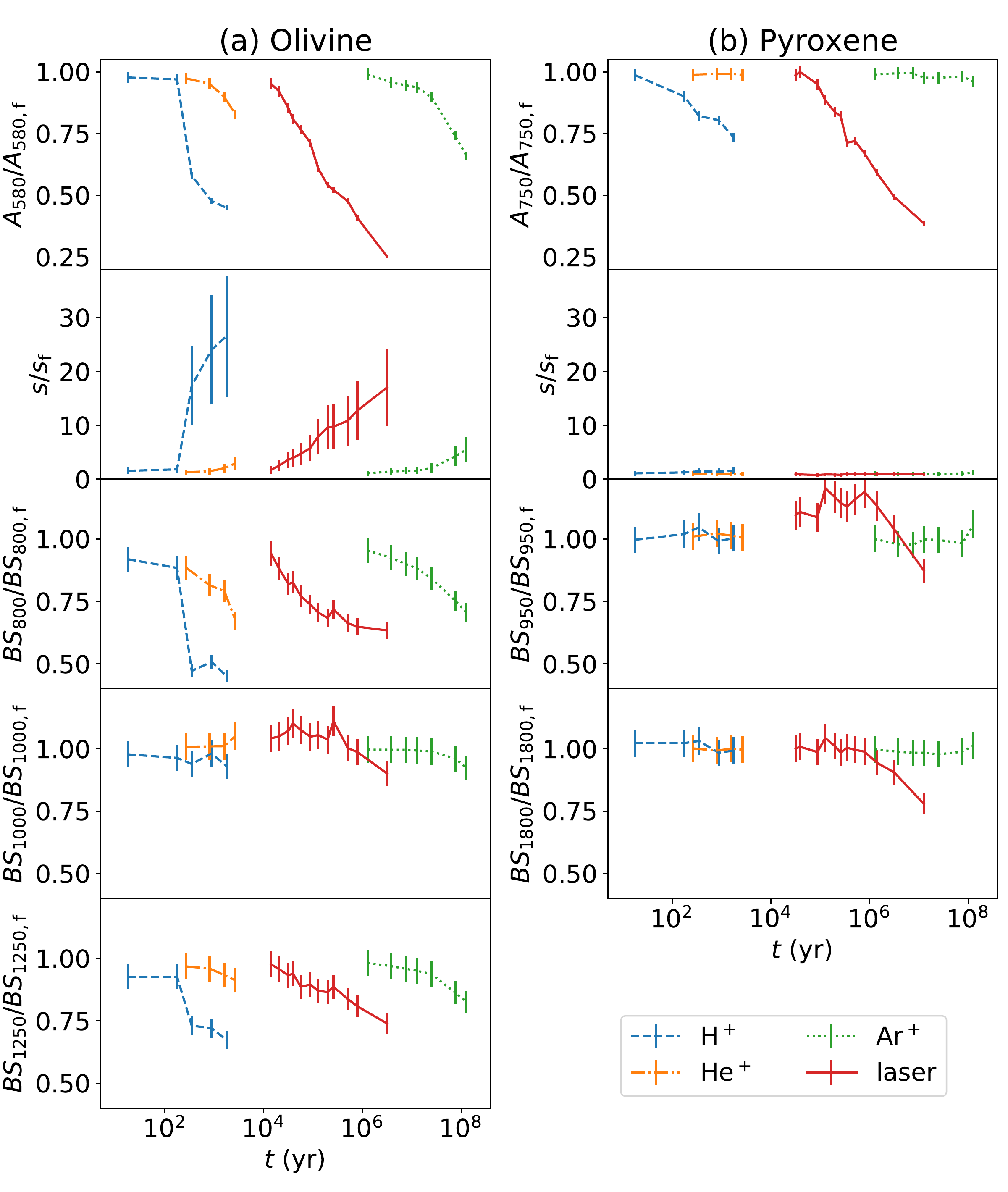}}
        \caption{Time evolution of spectral parameters of (a) olivine and (b) pyroxene. Each curve corresponds to one type of irradiation and connects points denoting individual irradiation steps. H$^+$ stands for hydrogen, He$^+$ for helium, and Ar$^+$ for argon ions. Time, $t$, was estimated for 1~au (see Appendix \ref{appendix_times} for more detail), $A_\lambda$ denotes albedo at a given wavelength, $s$ stands for the spectral slope, and $BS_\lambda$ stands for the strength of a band centred at a given wavelength $\lambda$. All parameters are divided by the corresponding spectral parameter of the fresh (f) material and include error bars. For details of pyroxene spectral slope changes, see Fig.~\ref{energy_evolution_fig}.}
        \label{time_evolution_fig}
\end{figure}

We fitted all of the measured spectra using the MGM, as described in Sect.~\ref{fitting_sect}, and plotted the spectral parameters as a function of astrophysical timescale, which we calculated from irradiation fluences and crater densities in accordance with the solar wind and dust fluxes at 1~au (see Appendix~\ref{appendix_times}). The evolutions of albedos, spectral slopes, and band strengths for each of the materials can be found in Fig.~\ref{time_evolution_fig}. We observe that even though the changes in albedo are of a similar order for olivine and pyroxene, pyroxene's spectral slope does not change much compared with that of olivine. 

We also note that the central olivine band does not change its strength significantly compared with the two outer bands (at $\approx$ 850 and 1250 nm). The band at $\approx$ 850 nm changed more (by $\approx\,55$\% for maximum irradiation) than the band at $\approx$ 1250~nm (30\%) (see Fig.~\ref{time_evolution_fig}).

\begin{figure}
        \resizebox{\hsize}{!}{\includegraphics{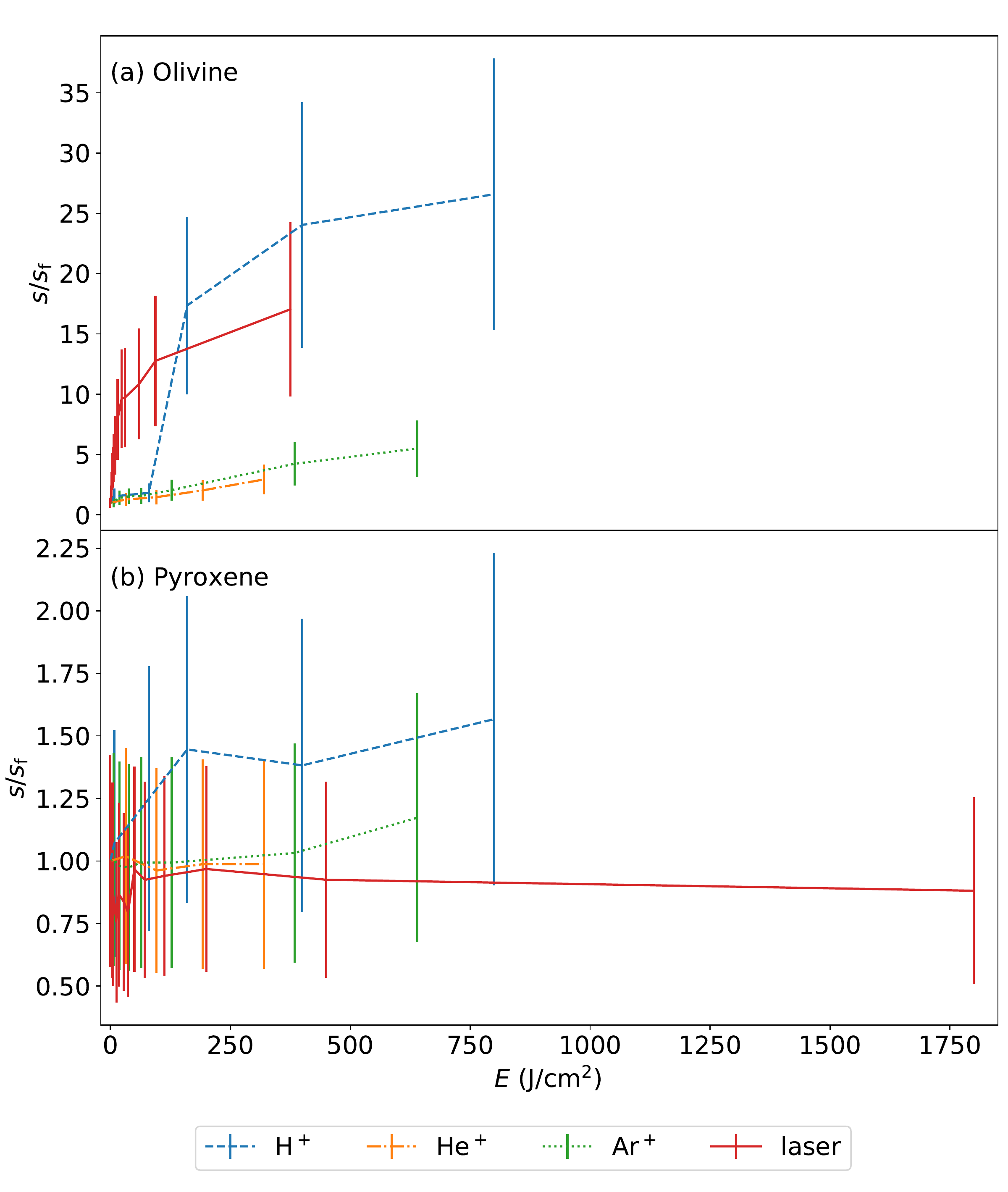}}
        \caption{Evolution of spectral slopes, $s$, of (a) olivine and (b) pyroxene with respect to the energy density, $E$. Each curve corresponds to one type of irradiation and connects points with increasing irradiation. H$^+$ stands for hydrogen, He$^+$ for helium, and Ar$^+$ for argon ions. Spectral slopes are divided by the corresponding spectral slope of the fresh material, $s_\mathrm{f}$, and include error bars.}
        \label{energy_evolution_fig}
\end{figure}

An example of how spectral parameters evolve with respect to the energy density delivered to the surface can be found in Fig.~\ref{energy_evolution_fig}, where the evolution of the spectral slope is shown. For olivine we can see that  samples irradiated by all ions follow a similar evolution up to a threshold energy density. Above this the H$^+$-irradiated sample undergoes a rapid change in the value of the spectral slope. This happens even for other studied parameters (albedo, band strengths, area of bands), but only in the case of olivine. Pyroxene does not show this behaviour in any of the parameters. See Appendix \ref{appendix_amorphisation} for more information and discussion.

We also find that laser influences the spectral slope of olivine at low densities of energy more than ions. However, comparing the evolution of the spectral slope from Fig.~\ref{energy_evolution_fig} to that in  Fig.~\ref{time_evolution_fig}, we see that the time needed to accumulate equivalent spectral change by laser or micrometeoroid impacts is several orders of magnitude greater than that needed in the case of H$^+$ irradiation at 1~au. 

\subsection{Short versus long wavelengths}
\begin{figure}
        \resizebox{\hsize}{!}{\includegraphics{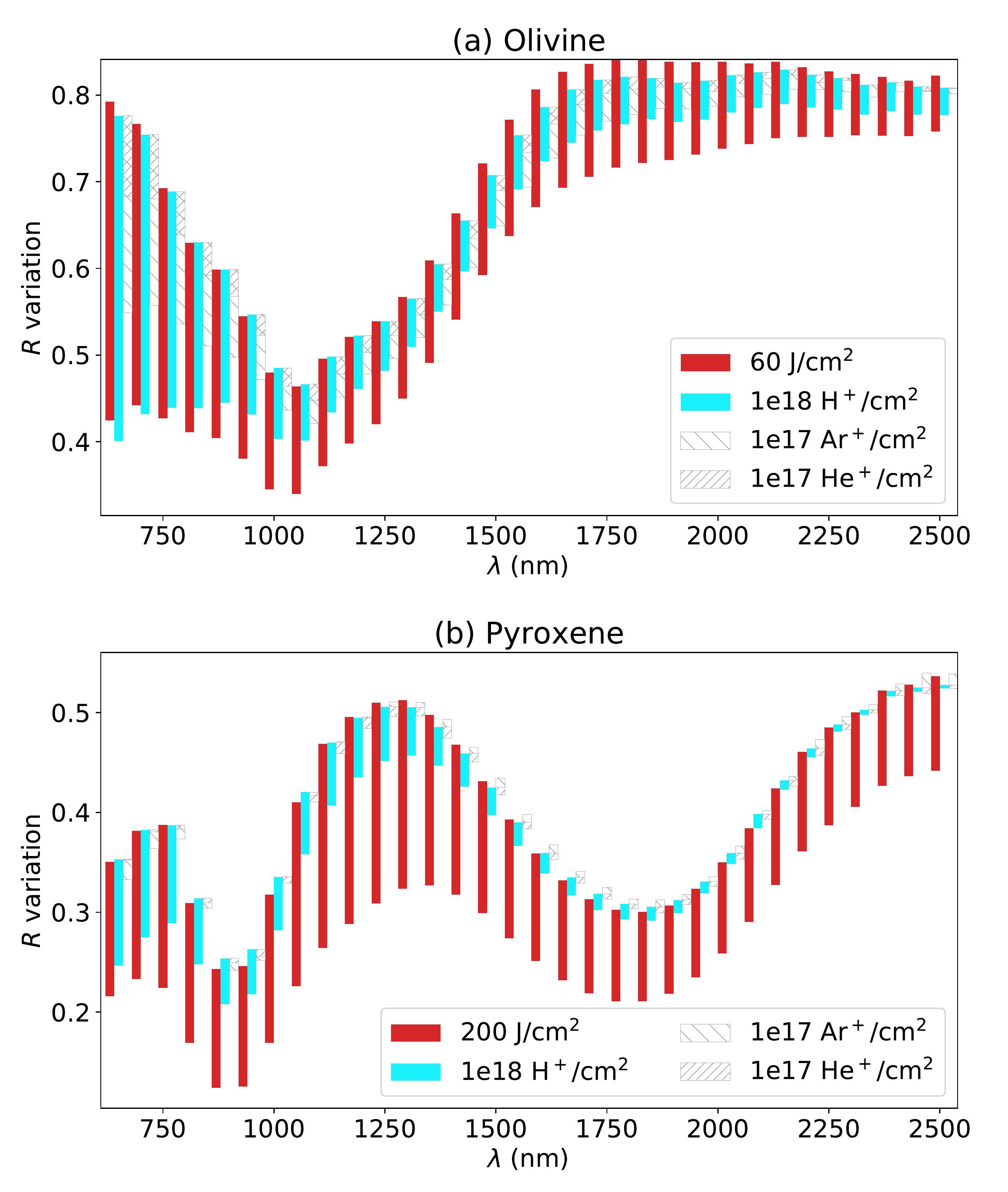}}
        \caption{Bar plot showing reflectance, $R$, variation induced by laser and ion irradiation experiments. Each bar connects at the upper end the reflectance of the fresh material and at the lower end the reflectance of the material weathered to the level stated in the legend. The length of the bar thus shows the variation in the reflectance at the given wavelength, $\lambda$. Instead of showing bars at all measured wavelengths, we down-sampled the spectra to 50 nm intervals to enhance readability.  H$^+$ stands for hydrogen, He$^+$ for helium, and Ar$^+$ for argon ions.}
        \label{bar_plot_fig}
\end{figure}

In Fig.~\ref{bar_plot_fig} we demonstrate the spectral differences related to the effects induced by laser and ion irradiation. Four different types of bars corresponding to different types of irradiation are plotted. The upper end of each bar indicates the reflectance of the fresh spectrum at the given wavelength. The lower end of the bar shows the reflectance of the spectrum irradiated by the fluence written in the legend. The longer the bar, the more the material changed due to ion or laser irradiation; in other words, the greater  the absolute change of the spectrum at that wavelength. 

We   intentionally selected the spectral curves of the laser and H$^+$ ion irradiation that have similar reflectance variation at VIS wavelengths, and studied how  the reflectance at NIR wavelengths behaves. We can see that the laser irradiation causes variations at long wavelengths greater than the H$^+$ ions, as seen, for example, from the ratio of the laser to the H$^+$ bar lengths at short and long wavelengths, which drops by 50\% in the case of olivine and by 70\% in the case of pyroxene. He$^+$ and Ar$^+$ ions do not show variations at the VIS wavelengths that are  as prominent  as those for H$^+$ ions because we did not reach high enough irradiation fluences. Nevertheless, the trend of decreasing reflectance variation towards longer NIR wavelengths is visible even for these two types of ions. Relative to laser bars, He$^+$ and Ar$^+$  bars drop by 23 and 43\% at longer wavelengths in olivine. In pyroxene Ar$^+$ shows a 24\% drop, and the change in  He$^+$  is insignificant.  

\begin{figure}
        \resizebox{\hsize}{!}{\includegraphics{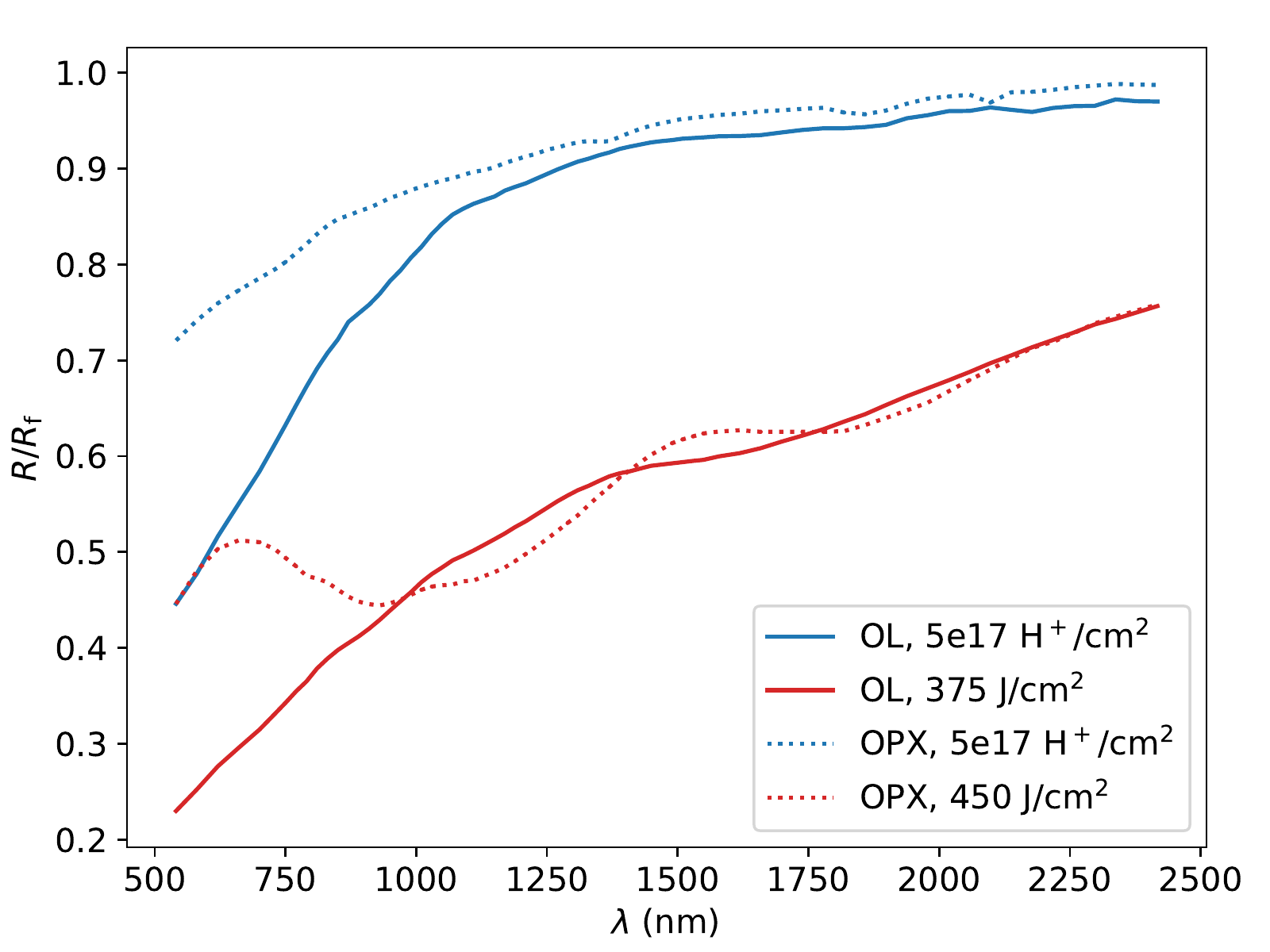}}
        \caption{Ratios of reflectance, $R$, of irradiated and fresh (f) spectra of olivine (OL) and pyroxene (OPX), irradiated by hydrogen (H$^+$) ions and laser. $\lambda$ stands for wavelength. A similar plot for mid-infrared measurements is in Fig.~\ref{appendix_mir_fig}.}
        \label{ratio_comparison_fig}
\end{figure}

In contrast to Fig.~\ref{bar_plot_fig}, Fig.~\ref{ratio_comparison_fig} shows relative changes in the spectra. It displays ratios of fresh to irradiated spectra. In this figure we plot spectra that represent similar energy density irradiation. There is a clear difference between olivine and pyroxene irradiated by laser. Olivine shows a major slope change in the VIS wavelengths, whereas in the case of pyroxene the slope change is not as pronounced, and we see that for the laser-irradiated pyroxene the absorption bands are altered. Furthermore, in the case of olivine, both types of irradiation cause much steeper changes at shorter than at longer wavelengths. The wavelength at which the behaviour changes is around 1 \textmu m.

\subsection{Albedo--slope correlation and principal component analysis}
\begin{figure}
        \resizebox{\hsize}{!}{\includegraphics{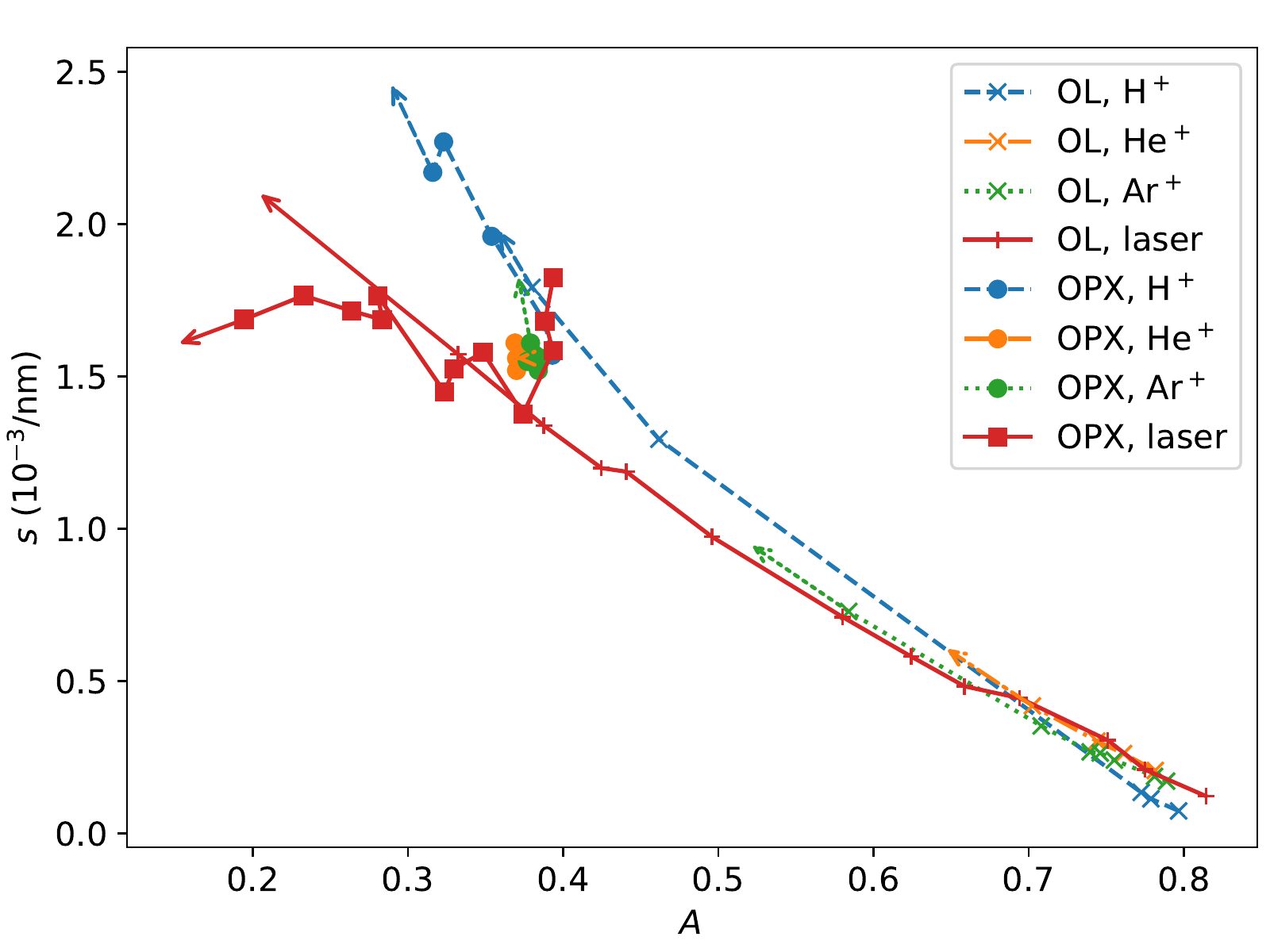}}
        \caption{Correlation of the albedo ($A$) and spectral slope ($s$) for olivine (OL) and pyroxene (OPX), subject to different irradiation set-ups. Each line connects the individual irradiated steps due to one type of irradiation. H$^+$ stands for hydrogen, He$^+$ for helium, and Ar$^+$ for argon ions. The highest irradiation states are indicated by the arrows' tips.}
        \label{ala_gaffey_fig}
\end{figure}

Similarly to \cite{gaffey_10}, we plotted a correlation of albedo (reflectance at a given wavelength) and spectral slope for our measurements (see Fig.~\ref{ala_gaffey_fig}). Each line corresponds to the evolution due to a different type of irradiation and markers on it highlight the individual irradiation steps. We see a clear trend of decreasing albedo and increasing spectral slope in all cases, except for the pyroxene irradiated by laser, for which mainly the albedo changes, but the spectral slope varies only slightly.

\begin{figure}
        \resizebox{\hsize}{!}{\includegraphics{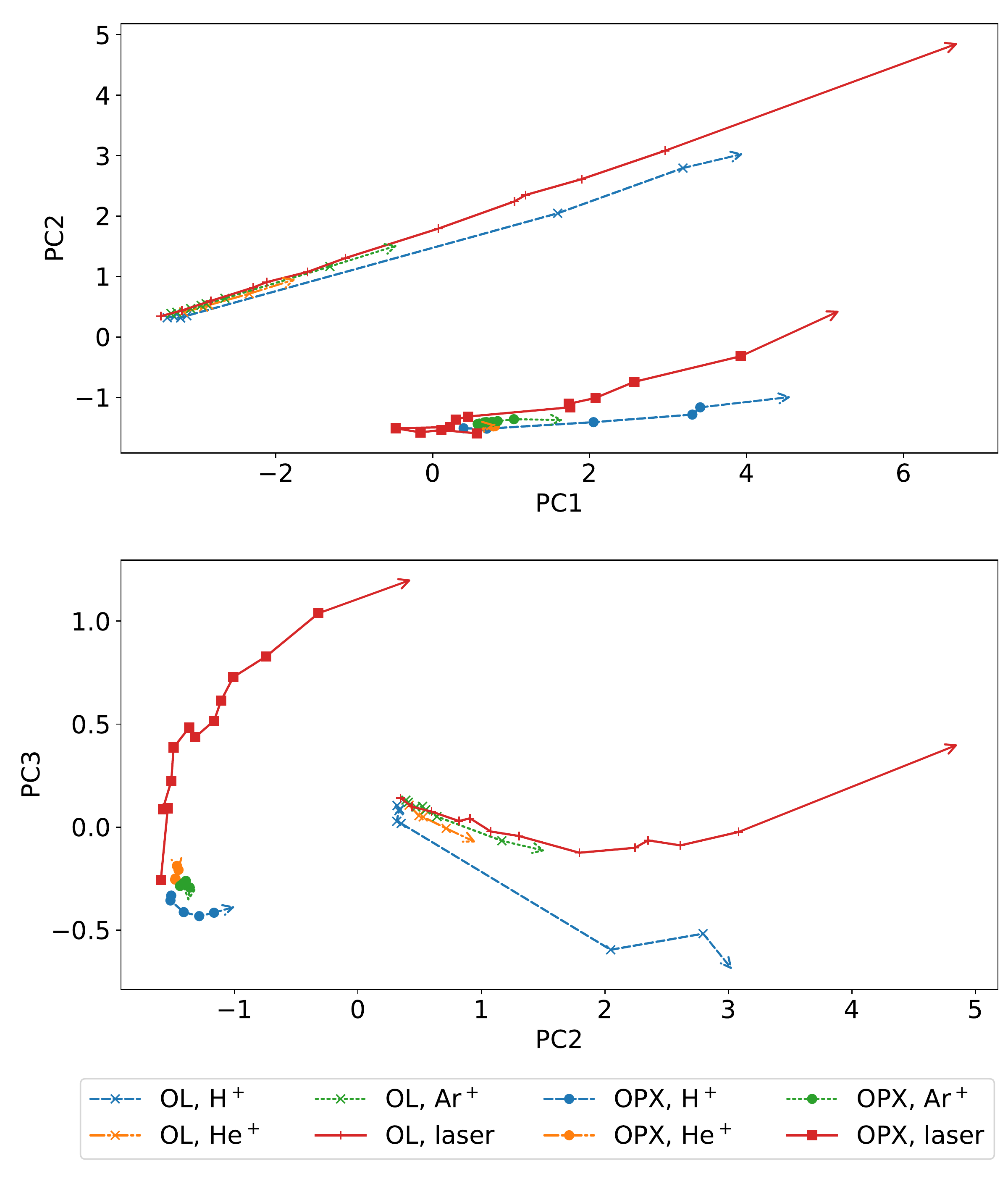}}
        \caption{Principal component (PC) plots for olivine (OL) and pyroxene (OPX) subject to different irradiation set-ups. Each point in the graph represents one irradiation state. H$^+$ stands for hydrogen, He$^+$ for helium, and Ar$^+$ for argon ions. The highest irradiation states are indicated by the arrows' tips.}
        \label{pca_fig}
\end{figure}

Additionally, the inclination of the curves is ordered H$^+$$\rightarrow$He$^+$$\rightarrow$Ar$^+$$\rightarrow$laser, thus from the lighter to the heavier ions, and then to micrometeoroid impacts. A similar sequence is seen in the principal component space (see Fig.~\ref{pca_fig}). We can clearly see that all the irradiation types follow approximately the same direction in the PC1 versus PC2 plot. The direction of the trends thus represents the space weathering evolution. From the first three, mainly the second principal component distinguishes between the different minerals. The separation of individual irradiation types is even more pronounced in the third principal component. 

\begin{figure}
        \resizebox{\hsize}{!}{\includegraphics{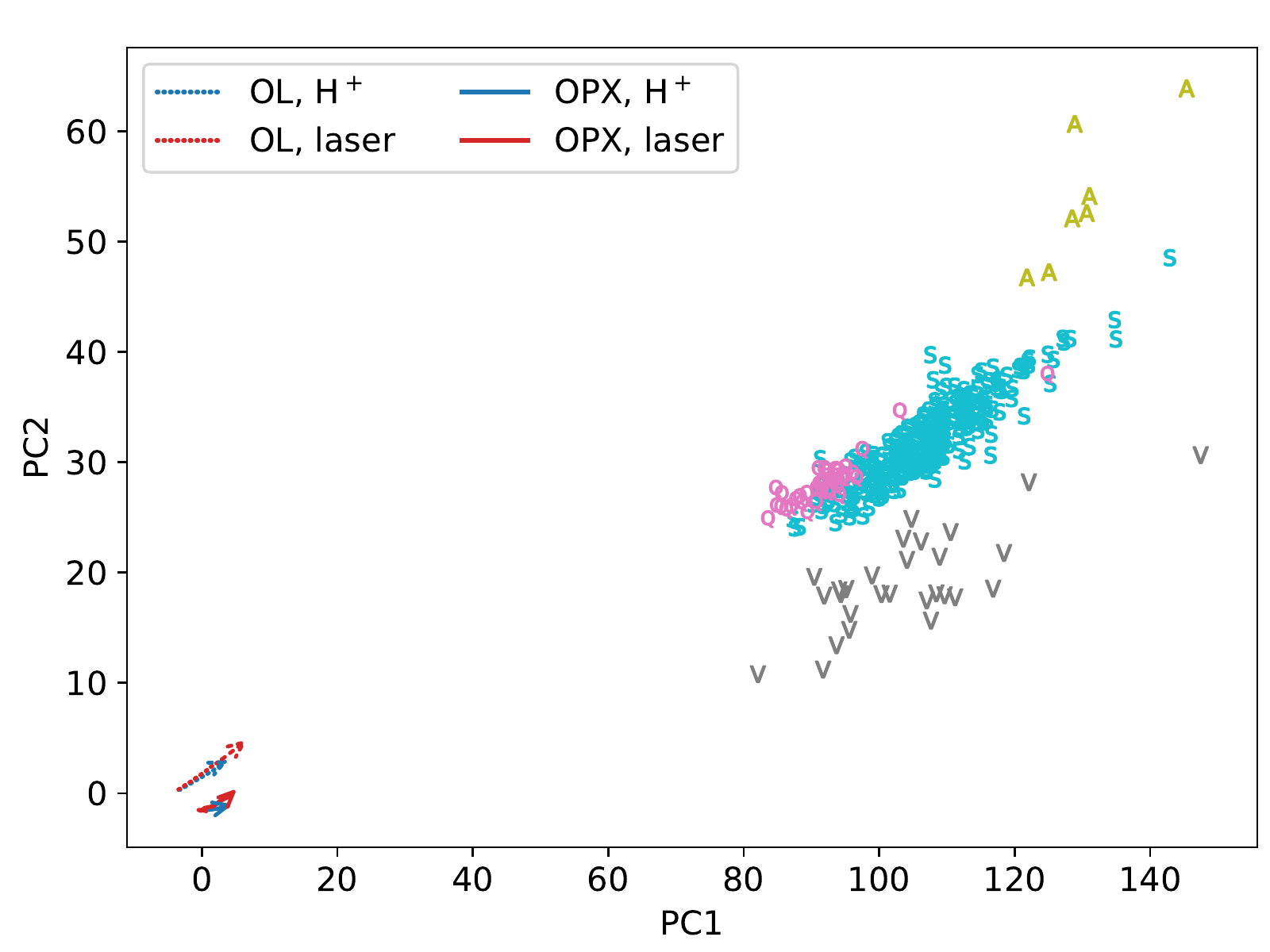}}
        \caption{Principal component (PC) plot for olivine (OL) and pyroxene (OPX) subject to different irradiation set-ups and including S-, Q-, V-, and A-type asteroids. Each letter (spectral type) in the graph represents the position of one asteroid in the principal component space. The spectra of irradiated pellets have the same positions as in Fig.~\ref{pca_fig} (top); for better readability, only hydrogen (H$^+$) and laser irradiation are shown.  The most irradiated spectra are indicated by the arrows' tips (the maturation thus goes from the lower left to the top right corner).}
        \label{pca-asteroids_fig}
\end{figure}

As described in Sect. \ref{pca_methods}, we also compared the asteroid spectra with our dataset (see Fig.~\ref{pca-asteroids_fig}). Even though the asteroids occupy a different part of the principal component space than our laboratory spectra, presumably due to the different origins, their distribution is significantly ordered. We find that the direction of weathering indicated by our experiments is parallel to the cloud of the Q- and S-type asteroids, with the Q-type asteroids at one end (in the less weathered direction) and the S-type asteroids at the other end. A- and V-type asteroids, as representatives of nearly pure olivine- and pyroxene-rich bodies, respectively, clustered into the separate clouds, and their orientation in the principal component space matches that of our laboratory spectra. 

\subsection{Mid-infrared spectra}
\label{mir_results}
\begin{figure*}
        \resizebox{\hsize}{!}{\includegraphics{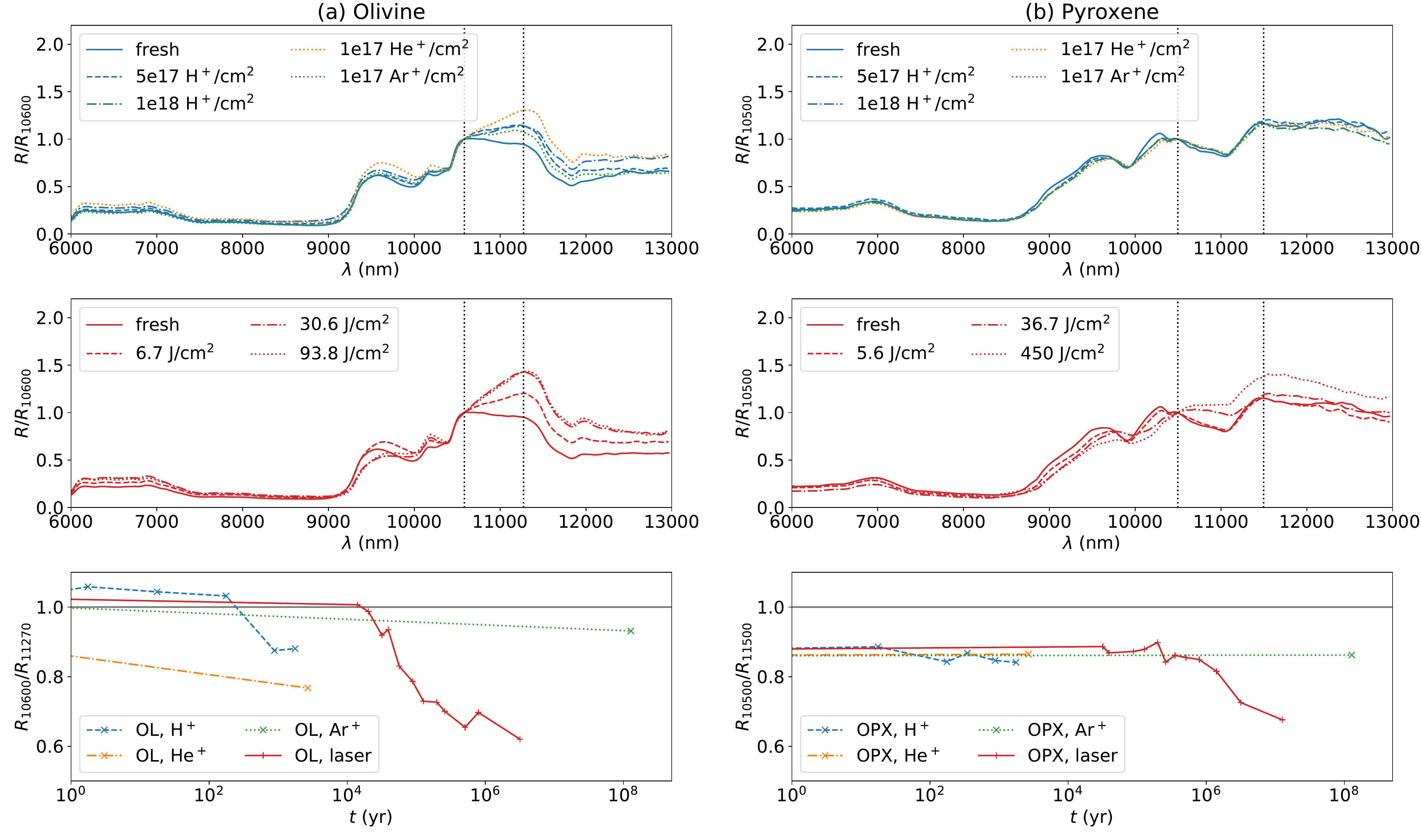}}
        \caption{(Top and middle) Mid-infrared part of some of the spectra of (a) olivine (OL) and (b) pyroxene (OPX) irradiated by ions and by laser. The two dotted lines show the approximate positions of the reflectance values ($R$), based on which  the ratio of the peaks was evaluated. The spectra were normalized at the wavelength ($\lambda$) of the first peak used for the ratio calculation to highlight the spectral alteration. (Bottom) Evolution of the reflectance ratio due to different irradiation set-ups. Time, $t$, was evaluated for the situation at 1~au (see Appendix~\ref{appendix_times}). A grey horizontal line indicates the situation of the flattened peaks. H$^+$ stands for hydrogen, He$^+$ for helium, and Ar$^+$ for argon ions.}
        \label{ol-plato_ratio_fig}
\end{figure*}

Figure~\ref{ol-plato_ratio_fig} shows selected olivine and pyroxene MIR spectra. We observe slight shifts in the positions of the reststrahlen bands (between $\approx$\,9.5 and 12 \textmu m) and also  alterations in the band's shapes, as also demonstrated in Fig.~\ref{appendix_mir_fig} (MIR wavelength range alternative to Fig.~\ref{ratio_comparison_fig}). We studied the evolution of the position of the Christiansen feature and found that the change is only several percentage points in our set-up. The most prominent change happened in both cases, for olivine and pyroxene, for the laser-irradiated samples. In the case of olivine, the Christiansen feature shifted to the longer wavelengths (from 8803.4 to 8986.5~nm), and for pyroxene to the shorter ones (from 8472.6 to 8191.5~nm for the maximum laser irradiation).

In Fig.~\ref{ol-plato_ratio_fig} (bottom) it is also apparent that the two major olivine reststrahlen bands at around 11 \textmu m changed their relative intensities with increasing irradiation. All fresh olivine spectra have a shorter-wavelength peak that is more intense than the longer-wavelength peak, and as the irradiation continues, the shorter-wavelength peak becomes less intense than the longer-wavelength one. This happens  for ion- and for laser-irradiated olivine. A similar conclusion can be made for pyroxene irradiated by laser. In the case of ion-irradiated pyroxene, the ratio  did not change very much.

\section{Discussion}
\subsection{Comparison with previous visible and near-infrared works}
As seen in the time and energy density evolution plots, Figs.~\ref{time_evolution_fig} and \ref{energy_evolution_fig}, spectral changes induced by different types of ions within one type of material are, in general, similar but they differ in their details. The reason for this is the different interaction of the incoming ion with the target material. The way an ion loses its energy in the material depends on the size of the ion, on its energy, and also on the target material \citep{brunetto_05}. We used three different types of ions, with two different materials. All of the experiments thus simulated different energy loss regimes. When an ion loses energy through inelastic (i.e. electronic) processes, the target material is subject to ionization. If the interaction is elastic (i.e. nuclear), the material is subject to the creation of vacancies, phonons, and recoils.

We calculated numerical simulations of our experiments using the SRIM software \citep{ziegler_85} to obtain information on the different regimes of elastic and inelastic energy losses in the material. An overview can be found in Table~\ref{srim_table}. We can see that the energy loss proceeds through inelastic collisions in the case of our H$^+$ and He$^+$ irradiation, as the electronic stopping power is larger than the nuclear stopping power. Ar$^+$ irradiation, on the other hand, is mostly subject to elastic collisions and produces the most vacancies per incoming ion of all the irradiation set-ups. We also calculated the SRIM simulations  for some  earlier published works to be able to compare their results with ours (see Table~\ref{srim_table}).
\begin{table*}
        \caption{Results of the SRIM simulations of the individual ion irradiation.} 
        \label{srim_table} 
        \centering
        \begin{tabular}{c c c c r c c c}
                \hline\hline
                & Mineral & Ion & $E$ (keV) & $d$ (nm) & $S_e$ (MeV/(mg/cm$^2$)) & $S_n$ (MeV/(mg/cm$^2$)) & Vacancies/ion\\
                \hline
                & & H$^+$       & 5     & 58 $\pm$ 29 & 0.204 & 0.007 & 3.4\\ 
                This work & OL  & He$^+$        & 20    & 132 $\pm$ 51 & 0.380 & 0.035 & 52.2\\
                & & Ar$^+$      & 40    & 31 $\pm$ 10 & 0.929  & 2.731 & 393.1\\
                \hline 
                & & H$^+$       & 5     & 61 $\pm$ 32 & 0.193 & 0.007 & 3.7\\
                This work & OPX & He$^+$        & 20    & 142 $\pm$ 57 & 0.352 & 0.034 & 56.8\\
                & & Ar$^+$      & 40    & 33 $\pm$ 11 & 0.850 & 2.644 & 414.8\\
                \hline
                \cite{loeffler_09} & & He$^+$ & 4 & 29 $\pm$ 19 & 0.017 & 0.073 & 21.1\\
                \cite{lantz_17} & OL & He$^+$ & 40 & 241 $\pm$ 70 & 0.530 & 0.022 & 66.1\\
                \cite{kanuchova_10} & & Ar$^+$ & 400 & 97 $\pm$ 62 & 2.544 & 1.214 & 2146.5\\
                \hline
                \cite{marchi_05} & OPX & Ar$^+$ & 200 & 152 $\pm$ 29 & 1.916 & 1.740 & 1441.2\\
                \cite{lantz_17} & CPX & He$^+$ & 40 & 252 $\pm$ 77 & 0.499 & 0.022 & 74.8\\
        \end{tabular}
        \tablefoot{Results of olivine (OL) and pyroxene (OPX) simulations. $E$ stands for the energy of the incoming ion; $d$ is the penetration depth; $S_e$ and $S_n$ are the electronic and nuclear stopping powers; H$^+$, He$^+$, and Ar$^+$ denote hydrogen, helium, and argon ions, respectively. In the model we used the following densities: 3.2~g/cm$^3$ for OPX and 3.3~g/cm$^3$ for OL. We also calculated SRIM simulations for some of the previous experimental work available in the literature, using compositions and densities relevant to the given experiment. CPX represents clinopyroxene.}
\end{table*}

\cite{loeffler_09}, who used olivine with similar forsterite content, obtained albedo changes very similar to those we see in our He$^+$ irradiation. As we can see from Table~\ref{srim_table}, Loeffler's set-up caused half the number of vacancies per ion, but the nuclear stopping power, which is, according to \cite{brunetto_14}, correlated with the spectral change, was twice as large. A larger albedo variation in samples irradiated by \cite{lantz_17} is probably connected to the higher energy of impacting ions, which caused  a  penetration depth of the ions that was two times greater (see the next section for more details on the connection of the penetration depth of ions to the spectral response). Comparing our Ar$^+$ irradiation with that of \cite{kanuchova_10}, we obtained   albedo changes that were twice as large,  but applied twice the energy per cm$^2$. This also corresponds to the values of the stopping powers in the table.

In the case of He$^+$ and Ar$^+$  irradiation, the  pyroxene spectra did not alter much. In contrast to our results, other work using pyroxene of a different composition achieved noticeable spectral change. For example, \cite{marchi_05} irradiated Bamble orthopyroxene using Ar$^+$ ions. Bamble orthopyroxene has an $\approx$\,20\% higher enstatite number than our samples \citep{bowey_20, cloutis_86}, which could have influenced the inclination of the sample to the space weathering related changes. Based on the SRIM simulation of Marchi's experiment the electronic and nuclear stopping powers are comparable, whereas in our case the nuclear stopping power was greater than the electronic stopping power (see Table~\ref{srim_table}). In addition, our configuration caused several times fewer vacancies per ion, which may have resulted in the different spectral evolutions as the material was less affected in our experiment.  

If we compare the spectral changes of our laser irradiation experiments with those in the literature, we  find alterations of a similar level, even though these were done mostly using nanosecond laser pulses that induce different post-irradiation processes in the sample. The olivine albedo changes are comparable for instance to those in the work of \cite{yazhou_17}, who used the same olivine material, and the pyroxene changes are similar to the work of \cite{sasaki_02}.

No matter what irradiation type we used, the central olivine band (at $\approx\,1000$~nm) changed its strength  less than the two outer olivine bands (at $\approx\,800$ and 1250 nm). The same behaviour has already been described by \cite{han_20}, among others. There is a theoretical explanation of this given in \cite{penttila_20}. The absorption in material follows approximately the Beer--Lambert exponential law. The same amount of darkening agent from space weathering in an initially brighter material will decrease the reflectance more compared with an initially darker material. The central olivine band (at 1000 nm) is the deepest in the fresh material, and thus will have the smallest absolute decrease. Additionally, the 800 nm band changed more significantly than the 1250 nm band, which correlates with the asymmetry of the imaginary part of the refractive index of olivine observed by \cite{brunetto_07b}.

\subsection{Wavelength-dependent changes and penetration depth}
\label{pen_depth_sect}
In Fig.~\ref{bar_plot_fig} we see that all of the ions influence the longer NIR wavelengths less than laser irradiation, while the impact of the laser and H$^+$ irradiation on shorter wavelengths was comparable. This results in more significant reddening of the H$^+$-irradiated compared with the laser-irradiated spectra in the time evolution plots (Fig.~\ref{time_evolution_fig}). 

An explanation for this difference can be found in the different penetration depths of the ions and laser (different mechanisms of the radiation damage). We used the SRIM software to evaluate the penetration depths of H$^+$, He$^+$, and Ar$^+$ ions in olivine and pyroxene. We obtained penetration depths ranging from 31 to 142~nm (see Table~\ref{srim_table}). Ion irradiation creates in situ damage of the surface layer, which happens over a relatively long timescale of the irradiation experiment and presumably causes in situ amorphization, crystalline lattice dislocations, and occurrence of nanophase Fe particles \citep{laczniak_21}.

Based on simple light scattering simulations using olivine and pyroxene slabs of different widths and using the extinction coefficients from the Database of Optical Constant for Cosmic Dust\footnote{\url{https://www.astro.uni-jena.de/Laboratory/OCDB/crsilicates.html#E} \citep{zeidler_11,huffman_71}.}, we estimated the penetration depth of the laser beam as $\approx100$ \textmu m. The timescale of the energy delivery is comparable to the timescale of the laser pulse ($\approx$\,100~fs) causing instant ablation of the irradiated volume. The ablated material is subsequently redeposited on top of the in situ damaged zone creating amorphous redeposition layer with possible presence of nanophase Fe particles \citep{weber_20}. Thus, the resulting pattern of the affected surface grains is expected to be more complex than in the case of ion irradiation. 

Based on the above-mentioned estimates, we can thus conclude that the laser penetrated much deeper into the sample than the ions. We can make a similar estimate for the penetration depths of the spectral beam. For olivine the penetration depth in the wavelength range of 0.7 to 2.5~\textmu m (VIS-NIR) is of the order of 100~\textmu m, while at 10~\textmu m (MIR region) it drops to $<$ 1~\textmu m. For pyroxene, the extinction coefficients are larger, and the penetration depths would thus be even smaller. The VIS and NIR spectral beams therefore work in the volume scattering regime, as commented also by \cite{brunetto_20}. Based on these numbers, the NIR spectral beam probes far deeper into the material ($\approx\,$100~\textmu m) than the depth of the material altered by ions ($<$~200~nm). The spectra thus represent a mixture of the fresh and irradiated material, and the spectral change is small. On the other hand, reflectance from the laser-irradiated pellet is influenced mainly by the altered material (because the penetration depth of the laser pulse is slightly greater than or similar to the penetration depth of the spectral beam); laser irradiation thus creates a greater variation in the NIR wavelength range. 

Additionally, as \cite{brunetto_07b} showed, after laser ablation the extinction coefficients of both olivine and pyroxene increase throughout the VIS and NIR wavelengths, which results in a decreased penetration depth of the light. The change in the coefficient is orders of magnitude larger for the VIS wavelengths than for the NIR. \cite{penttila_20} modelled evolution of olivine extinction coefficients and found that inclusions of nanophase iron alter the VIS extinction coefficients $\approx$~20 times more than the extinction coefficients in the NIR wavelength range. The reason for this is the different size of nanophase iron particles relative to the wavelengths of the VIS and NIR light. Assuming that  changes such as \cite{brunetto_07b} and \cite{penttila_20} found are similar for our ion irradiations, we can explain why in the VIS wavelengths the reflectance variation in H$^+$- and laser-irradiated samples is comparable:  the relative contribution of the H$^+$-irradiated layer is probably larger in the VIS than in the NIR range. The MIR region is subject to the surface scattering regime, where all the four types of irradiation caused damage, and thus would be indiscernible. 

Another way to visualize the different behaviour at the short and long wavelengths of our spectra is shown in Fig.~\ref{ratio_comparison_fig}. Especially in the case of olivine, a significant change in the slope alteration at around 1\,000 nm is visible. This figure also shows that even though the ions mainly modify the spectral slope, pyroxene irradiated by laser displays very little slope change, and we observe mainly the band depth changes. 

Supporting our findings, \cite{weber_20} pointed out that depending on the space weathering agent, the micro-structure of the minerals in space is influenced significantly differently.

\subsection{Space weathering versus mineralogy}
Even though we observed a difference in the effect of ions and laser pulses at the NIR wavelengths, generally we observed that the initial mineralogy of the pellet has a more significant effect on the final evolution of the spectral features than the type of irradiation. 

In Fig.~\ref{time_evolution_fig} we see that even though the albedo  changes of olivine and pyroxene are of a similar order, they differ significantly when considering the evolution of other spectral parameters, especially the spectral slope. Olivine shows major changes in the values of the spectral slope, but pyroxene changes only mildly. We can also see this trend  in Fig.~\ref{ala_gaffey_fig}, where especially pyroxene irradiated by laser shows major changes in albedo but nearly no change in spectral slope.

Such inconsistency of behaviour was studied on an atomic scale by \cite{quadery_15}. When OPX is irradiated by ions, its crystalline structure suffers defects. \cite{quadery_15} showed that such defects are easily healed by ions from the bulk (subsurface layers) of the sample, which ensures that this material stays stable longer. On the other hand, olivine does not have such an ability, and thus weathers more quickly as there are no species that would renew the topmost layers.

A similar difference in susceptibility to the laser irradiation was reported by \cite{weber_20}. Based on their transmission electron microscopy analysis of olivine and pyroxene (both of similar composition to our samples), they found that the thickness and also the nanostratigraphy of the layer altered by the irradiation differs for these two minerals, which is consistent with the different penetration depths of light in olivine and pyroxene that we noted in Sect.~\ref{pen_depth_sect}. 

The PCA pointed to the space weathering trend in the measured data. All the irradiated steps, due to one type of irradiation, lie on approximately straight lines in the plot of the first two components in Fig.~\ref{pca_fig} (top). Remarkably, in each of the compositions, the H$^+$-irradiated evolution lies at the lowest values of PC2 and the laser-irradiated case at the highest PC2 (the same holds for Fig.~\ref{pca_fig}, bottom). More significantly, the PCA divides the measured spectra into two main clouds based on the composition. The PCA thus identifies the different mineralogies. Furthermore, the asteroid spectra also cluster  based on the mineralogy in the principal component space (Fig.~\ref{pca-asteroids_fig}) and their orientation matches that of our laboratory spectra. This means that our spectra are a suitable analogue to the asteroid data. The plot of the first two principal components shows the trends due to the influence of the micrometeoroid impacts and solar wind only as second-order information. Only the third principal component is capable of distinguishing some differences in various irradiation types.

\subsection{Consequences for asteroid studies}
\subsubsection{Pyroxene-rich asteroids}
As we discuss above, pyroxene experiences only minor changes in spectral slope, and the strengths of the absorption bands are also quite stable when compared to the amplitude of changes in olivine. On the other hand, H$^+$ ion and laser irradiation induces major changes in pyroxene albedo. 

We see such inconsistency in the amplitude of changes of different spectral parameters even in some asteroids, for example (4)~Vesta. As \cite{pieters_12} pointed out, Vesta's surface shows a range of pyroxene compositions. From observations of the Dawn mission, very strong absorption bands and nearly no continuum changes are seen, even though there are various dark and bright regions. The albedo   changes, although the absorption bands are strong and thus look fresh. \cite{pieters_12} claim that Vesta does not show the typical (lunar-type) space weathering caused by an accumulation of the nanophase Fe particles, but rather a small-scale mixing of diverse surface components. \cite{vernazza_06} tried to simulate space weathering on Vesta through ion irradiation of a eucrite meteorite (presumable analogue) and, unlike us, they found that the spectrum should change even in terms of the spectral slope. As the slope changes are not observed on Vesta, \cite{vernazza_06} concluded that Vesta's surface is shielded from the solar wind by a magnetic field. Instead, our pyroxene did not show significant reddening of the spectral slope; the weathering at Vesta thus may proceed even without the presence of contamination by carbonaceous material or a magnetic field. In addition, as \cite{fulvio_12} showed, there is variation in the slope behaviour of individual eucrite meteorites upon ion irradiation.

\subsubsection{Olivine-rich asteroids}
As shown in Fig.~\ref{time_evolution_fig}(a), the spectral slope of H$^+$-irradiated olivine pellets increased considerably over only several hundred  years (at 1 au, for the main asteroid belt, the value would be approximately 10 times longer). This means that the solar wind component of the space weathering has a significant and rapid effect on olivine-rich asteroids, explaining the observed high slopes of A-type asteroids (see e.g. \citealt{cruikshank_84} and \citealt{demeo_09}). For example, asteroid (1951) Lick shows a $10^7$ to $10^8$ year old surface, as \cite{brunetto_07} showed using their laser-ablated olivine samples together with a scattering model.

\subsubsection{Q- and S-type asteroids}
Q-type asteroids show spectra that are very similar to ordinary chondrite meteorites, thus fresh. Their spectra also show remarkable similarity to those of S-type asteroids. The difference is that the spectra of Q-type asteroids are brighter and have less steep slopes and deeper absorption bands than those of S-type asteroids. From earlier observations, Q-type asteroids are small and are mostly located in the near-Earth region. A transition between Q- and S-type asteroids has been observed for asteroid sizes between 0.1 and 5~km \citep{binzel_04}. The accepted theory is that the surfaces of Q-type asteroids look fresh because they rejuvenate more easily than those of S-type asteroids. The main reasons may be that as they are smaller, their collisional lifetimes are shorter, and also their presence in the near-Earth region makes the tidal effects, including seismic shaking, more significant \citep{vernazza_09}. 

In Fig.~\ref{pca-asteroids_fig} we can see that the Q-S distribution within the asteroid cloud is the same as the orientation of the fresh-irradiated distribution within both olivine and pyroxene samples. It is important to note  that the positions of the asteroids in the principal component space are based on information regarding our measured spectra (not on information from the whole asteroid dataset) and still they cluster in a manner that agrees with the Q- to S-type transition.

Our PCA simulation thus verifies that there is not a large compositional difference in the S- and Q-type asteroids, and the variation in their spectral features can be associated with the different weathering stages. On the other hand, the PCA did not give any clue to whether there is a difference in the dominant space weathering agent between the Q- and S-type asteroids, as the directions of the different simulations are nearly parallel to the Q--S cloud. Based on this technique we are therefore unable to judge the history of the space environment of the individual bodies; in other words, we cannot say whether the body was subject to an environment with increased micrometeoroid flux or whether the weathering stage was caused mainly by solar wind ions.

S-type asteroid (433)~Eros shows high albedo contrast ($\approx\,36$\%), but the slope variations are not as significant as we would expect during typical weathering circumstances \citep{clark_01,murchie_02}. An explanation may be that on the old surface of asteroid Eros, the spectral parameters of olivine are already saturated and the only variation present in the spectra is due to the changes in pyroxene, which take longer to propagate and are minor with respect to the spectral slope. Asteroid Itokawa, on the other hand, shows large spatial variation in the spectral slope. The spectra of Itokawa can in some parts of the surface be classified as Q-type and in other parts as S-type.  The changes in the spectral slope on this asteroid suggest that the olivine spectral signatures are not yet saturated, as in the case of Eros, and that Itokawa may represent a younger surface. Slopes of Itokawa's spectra, however, are evaluated based on information relying on the incompletely captured 1~\textmu m band \citep{abe_06, hiroi_06, koga_18}.

\subsection{Mid-infrared spectra}
\label{mir_discussion}
As mentioned in Sect.~\ref{mir_results}, the position of the Christiansen feature behaves inconsistently, shifting to the opposite direction, for olivine and pyroxene irradiated by laser. \cite{lucey_17} evaluated the variation in the position of the Christiansen feature on the Moon and concluded that the feature shifts to  longer wavelengths as the maturation process continues, which is in agreement with what we see in olivine and with what \cite{weber_20} observed, and is in contrast to the previous work by \cite{salisbury_97}, who found that the Christiansen feature is not much affected by changes induced by space weathering.

Shifts in the position of the Christiansen feature are probably correlated with the visual albedo of the material, as noted by \cite{lucey_17} and references therein. In a vacuum, the albedo influences the thermal gradient of the surface layer \citep{henderson_97}, which determines the depth from which the thermal emission originates and affects the Christiansen feature wavelength position (see also recent article by \citealt{lucey_21} for more details). However, albedo is probably not the only factor influencing the position of the Christiansen feature. For example, \cite{lantz_17} pointed out that preferential loss of magnesium to Fe during the maturation process shifts the positions of the bands to longer wavelengths. 

The possible reason for the inconsistency of the Christiansen feature shift between olivine and pyroxene may be the inappropriate spectral resolution. Different sampling of the spectral curves has an influence on the estimation of the Christiansen feature's position, as mentioned by \cite{salisbury_91}. Alternatively, a change in the position of the Christiansen feature is too small (of the order of a few per cent) to make a general conclusion based on it. 

In Fig.~\ref{ol-plato_ratio_fig} (bottom), we show that the relative intensities of the olivine reststrahlen bands at around 11~\textmu m change as the irradiation proceeds because the shorter-wavelength band is altered more significantly than the longer-wavelength band. The same holds for pyroxene irradiated by laser. Similar changes in the relative intensities of the reststrahlen bands have already been observed, for example, by \cite{brunetto_20}. This evolution may be attributed to changes in the structure of the material or to changes in the preferred orientation of the crystals (see e.g. \citealt{weber_20}).

We observed that the time needed to flatten the peaks of the restsrahlen bands differs for the different irradiation types. We should  note that there is a dependence of the position of the reststrahlen bands on the observation geometry, as was reported by \cite{rubino_20}. In order to compare the asteroid observations to the laboratory data, one thus needs to make sure that the observational set-ups are the same.

\subsection{Timescales}
As already mentioned, if we compare the total energy density input of various irradiation techniques with the spectral change (see Fig.~\ref{energy_evolution_fig}), at low energy densities laser causes greater changes than any of the ions. However, we did not evaluate the energy budget of individual processes that happen after irradiation and have the final effect on the spectra, so this result is only tentative. 

If we consider the time needed to accumulate the same energy density through micrometeoroid impacts and through solar wind irradiation at 1~au, it is significantly longer for micrometeoroids (see Fig.~\ref{time_evolution_fig}). Looking, for example, at the changes in olivine, a 50\% drop in albedo takes 1000 times longer by means of micrometeoroid impacts than by solar wind ions. This result is consistent with previous laboratory work, which showed that the typical timescales of the significant spectral change are different: for the solar wind irradiation $10^4$ to $10^6$ years, and for micrometeoroid bombardment $10^8$ to $10^9$ years \citep{sasaki_01a,noguchi_11,blewett_11}.

\section{Conclusions}
Our experiments revealed that the H$^+$ irradiation, simulating solar wind, and the laser pulses, simulating micrometeoroid impacts, induce spectral changes that are comparable in the VIS wavelength range, but there is a significant difference in their ability to influence longer NIR wavelengths. Laser pulses cause greater variation in the reflectance at longer wavelengths than any of the ions we used. This may be due to the greater penetration depth of the laser pulses when compared to the that of ions. For this reason, in the  case of ions the material probed by the spectrometer is a mixture of a thin altered layer and the underlying unaltered layers, and the spectrum is thus less modified. 

From the comparison of the total energy input of various irradiation techniques with the spectral change, the H$^+$ ions of the solar wind affect the surface on timescales three orders of magnitude shorter than the micrometeoroid impacts. We can thus make the following prediction: the spectrum of a fresh planetary surface will probably be altered in the first $10^5$ years  mainly by the solar wind. It will therefore darken and significantly change its spectral slope (due to relatively smaller spectral change at longer NIR wavelengths). As the contribution of the micrometeoroid impacts  increases for the older surfaces, the material will continue to darken, and as the long wavelengths are also  altered, the spectral slope will not be modified to such an extent. By studying the NIR wavelength range, we can thus draw conclusions on the surface exposure ages of various different geomorphological features in the solar system bodies.

Except for the above-mentioned difference between ion and laser irradiation timescales and slope behaviour, we did not find any major dissimilarities in their effect on spectral evolution. It seems that the initial mineralogy is the leading factor influencing how the spectra will be affected regardless of the space weathering agent. In other words, all   types of irradiation cause similar changes to the spectra of the given mineral, but for olivine, the evolution of the spectral curves is different to that of  pyroxene.

While olivine shows prominent rapid changes in all spectral parameters in the VIS--NIR range, in the case of our pyroxene samples, we observed a slower response with only minor changes of the spectral slope compared with the magnitude of change in olivine. This may further contribute to the different slope behaviour of olivine-pyroxene mixtures (for instance Q- and S-type asteroids) on different timescales. In early phases, weathering of olivine shows rapid and significant slope change together with darkening or 1~\textmu m band depth change. Once olivine matures, the continuation of space weathering affects mainly pyroxene with related change in the albedo rather than in the spectral slope.

In the MIR region the shifts in the Christiansen feature minimum were opposite for olivine and pyroxene. We have seen that the relative intensities of the olivine reststrahlen bands change irrespective of the space weathering agent. In pyroxene the changes in relative intensities were caused mainly by laser irradiation; ion irradiation caused only a minor alteration.

Our results have significant implications for asteroid studies. For instance, the rapid changes of olivine's spectral slope observed during the H$^+$ irradiation explain the absence of A-type asteroids with flat or low spectral slopes. Our PCA results agree with the Q- to S-type asteroid transition.

\begin{acknowledgements}
We thank the following for their help with preparing the experiments: Arto Luttinen for supporting us with the pyroxene material, Yazhou Yang for the olivine material, Janko Tri\v si\' c Ponce for helping with pellet preparations, Romain Maupin, Radoslaw Michallik, Joonas Wasiljeff, Anna Vymazalová, Roman Skála, Petr Mikysek, and Zuzana Korbelová for analyses leading to revelation of the contamination in olivine, and Ji\v rí Pavlů and Viktor Johánek for the loan of equipment from their laboratories. We are also grateful to Jessica Sunshine for the discussion of MGM fits. The authors are grateful to the anonymous referee for careful reading of the paper and valuable suggestions and comments.
        
This work was supported by the University of Helsinki Foundation and the Academy of Finland project nos 325805, 1335595 and 293975, and it was conducted with institutional support RVO 67985831 from the Institute of Geology of the Czech Academy of Sciences. The authors acknowledge funding from Charles University (Project Progres Q47). Part of the irradiation was performed using the INGMAR set-up, a joint IAS-CSNSM (Orsay, France) facility funded by the French Programme National de Planétologie (PNP), by the Faculté des Sciences d’Orsay, Université Paris-Sud (Attractivité 2012), by the French National Research Agency ANR (contract ANR-11-BS56-0026, OGRESSE), and by the P2IO LabEx (ANR-10-LABX-0038) in the framework Investissements d’Avenir (ANR-11-IDEX-0003-01).

\end{acknowledgements}

\bibliographystyle{aa} 
\bibliography{literature} 

\begin{appendix}
\section{Experimental set-up}   
\label{appendix_setup}  
Our spectral measurements were made using three different spectrometers, with varying observation geometries; in one case we used the integrating sphere, and in the other two cases we measured with fixed incidence and collection angles, in each case slightly differently (see Table~\ref{experimental_setup_table} for more details). We did not apply any corrections to the spectra to equalize the observation circumstances. There are two main reasons for this. Firstly,  in the majority of cases we were comparing spectra with their fresh analogue measured with the same spectrometer. The relative trends thus did not suffer from the lack of correction. Secondly, as \cite{beck_12} showed, in the case of the Moon the difference between the spectral slope measured at a phase angle of 15$^{\circ}$ and 30$^{\circ}$ (our situation) is only a few percentage points, which is within the limits of our error bars. In addition, \cite{sanchez_12} pointed out that in their sample of 12 Q- and S-type asteroids, there needs to be a change in phase angle of about 90$^{\circ}$ to obtain spectral changes similar to those of the space weathering effects.

\begin{figure}
        \resizebox{\hsize}{!}{\includegraphics{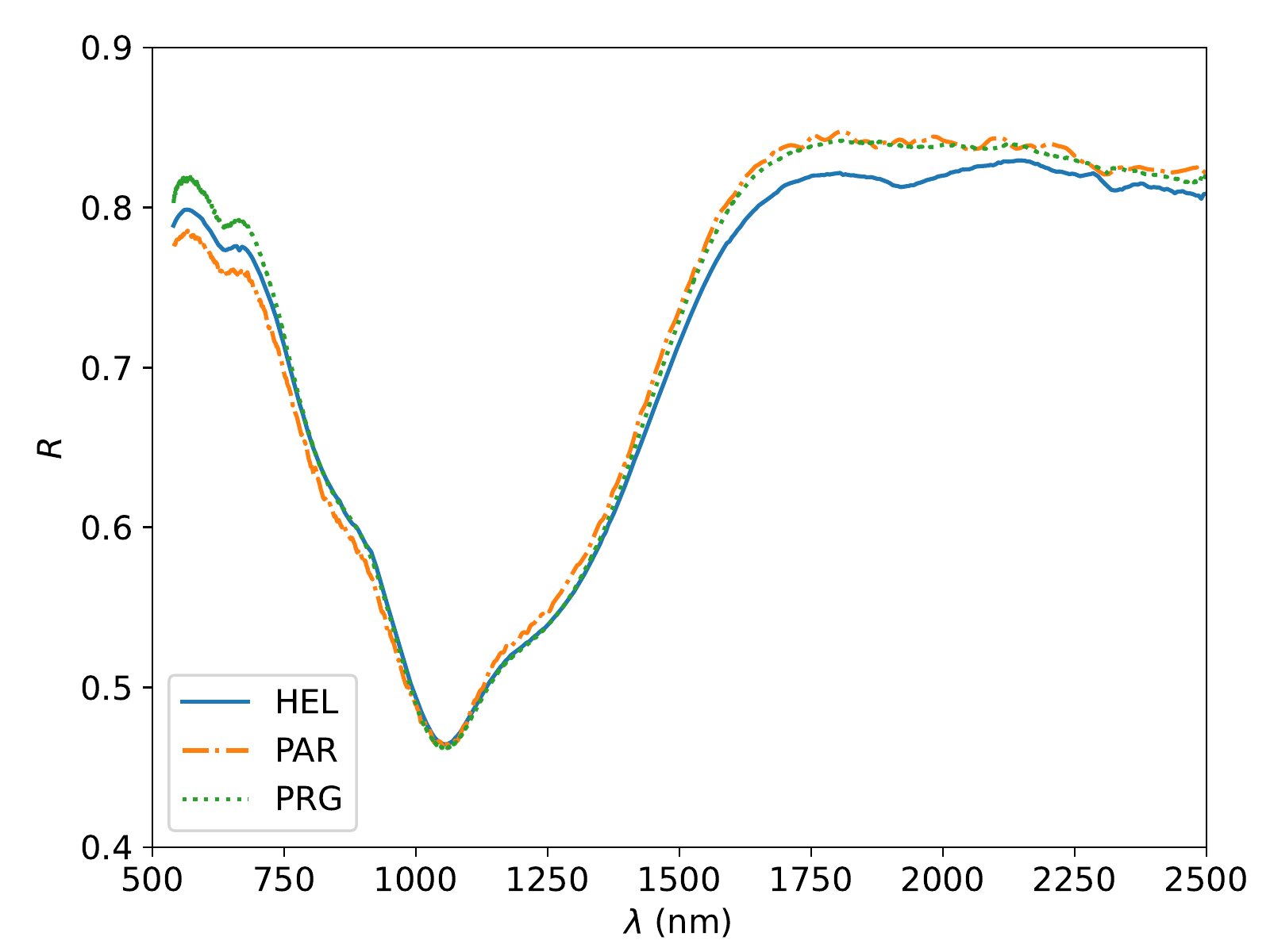}}
        \caption{Comparison of fresh olivine spectra measured on different spectrometers. Reflectance is marked by $R$, wavelength by $\lambda$, HEL indicates the University of Helsinki laboratories, PAR indicates the INGMAR set-up at IAS-CSNSM in Orsay, and PRG indicates the laboratories at Charles University in Prague.}
        \label{comparison_spectra_fig}
\end{figure}

Additionally, the integrating sphere spectrometer we used has a port in the sphere in the specular direction. The port can be closed with a PTFE-coated plate or opened with a beam trap attached to the port. In our spectral measurement we always used the configuration with the beam trap in the specular direction to avoid specular reflectance that would affect the measurements. A comparison of measurements of a fresh olivine pellet on different spectrometers can be seen in Fig.~\ref{comparison_spectra_fig}. We can see that the difference is indeed only a few percentage points.

Even though there were slight variations in the wavelength ranges of individual spectrometers and some spectrometers could provide information for a wider wavelength range, for the final evolution we used those spectral ranges that were covered by all the set-ups. For the VIS--NIR spectra we worked from 540 to 2\,500~nm and for the MIR spectra from 3 to 13 \textmu m.
        
\section{Notes on contamination}
\label{appendix_contamination}
The rock samples we used for the experiments were terrestrial minerals. Compared with the space environment, some contamination is possible even with strict experimental handling, which includes cleaning the equipment using ethanol or deionized water, wearing gloves, and covering the samples during processing to protect them from the airborne dust.

The OPX samples contained pargasite, as said in the main text, which we accounted for. Otherwise the OPX samples were pure, but in the case of olivine we saw some sort of contamination in the form of darker grains. Even in the MIR wavelength range, at around 3\,400~nm, we found several peaks. These peaks belong to organic material, most probably hydrocarbons and carboxylic acids, that may originate from terrestrial dust \citep{salisbury_91}. 

Analysing the samples using a scanning electron microscope (Tescan, Vega~3~XMU) with the energy dispersive spectroscopy method, we found that the majority of the contamination was a chromium-rich spinel (most probably chromite). Nevertheless, \cite{cloutis_04} and \cite{isaacson_10} showed that chromite is a common mineral in the solar system; it is usually found together with olivines in meteorites, in S- and K-type asteroids, and in lunar samples. 

Based on X-ray diffraction measurement, we conclude that the contamination is less than 4\,\% (limit of reliability of the measurement), and the contamination thus has only a minor effect on our results. To give more precise numbers, by a comparison with the uncontaminated sample we estimated that the absolute spectral difference between contaminated and uncontaminated material is less than 3\% in the area up to 650~nm and is even smaller at longer wavelengths. Among six fresh olivine pellets, the   albedo varies by 1.7\%, the strength of the absorption bands varies by less than 5\%, and the positions of the minima of the bands do not change. 

Chromite's spectrum is generally darker than the spectrum of olivine, and has absorptions around 2~\textmu m (see e.g. \citealt[Fig. 12]{cloutis_04}). We do not see any alteration of the olivine spectrum around 2~\textmu m, which is in contrast to \cite{cloutis_04}, who pointed out that spectral differences of olivine and chromite allow the detection of even small abundances of chromite in olivine. The effect of the contamination on the spectra is thus minor.  

\section{Timescales}
\label{appendix_times}
Considering the ion irradiation experiments, the following estimates of the astrophysical timescales are a proxy for the situation at 1~au. If  we wanted to evaluate the experiments, for example, in the main asteroid belt, the timescales would obviously be longer, as the flux of the solar wind ions would be smaller, as it decreases approximately with the square of the distance from the Sun \citep{schwenn_00}. We also use numbers valid for the typical situation in the solar system, which is the slow solar wind. In the case of solar wind flares, the fluxes and energies of the particles would be different. It should be  noted that compared with the real solar wind, the energies of our ions were slightly higher, which means that    our timescales should be taken  as order of magnitude estimates rather than exact values. On the other hand, the relative values belonging to one type of irradiation are valid.

To calculate what timescales the irradiation fluences of ions corresponds to, we need to compare the experimental values with the typical fluxes of the individual ions. Based on \cite{schwenn_00} the flux of protons at 1~au is $\approx\,2.9\times10^8$~ions/cm$^2$/s. If we consider the ratio of H$^+$ to He$^+$ ions in the solar wind, we get a flux of $1.9\times10^7$~ions/cm$^2$/s for He$^+$. For Ar$^+$, we have $4\times10^2$~ions/cm$^2$/s \citep{kanuchova_10}. By dividing our fluxes by the literature-based values,   in the first approximation we obtain the time that the material would be subject to irradiation at 1~au. To further improve the estimate, we approximate the object as a rotating sphere, in which case a correction factor of four is introduced into the calculation of the timescale. For bodies of other shapes the correction factor would be different, but this calculation is beyond the scope of our article and the relative trends we obtain in the first-approximation estimate are unaffected by the simplified solution.

Based on the number of individual pulses that were used during the laser irradiation, we can estimate the corresponding time the surface would experience a similar number of impacts at 1~au. The flux of dust particles several microns in size is, based on \cite{sasaki_02} or \citet[page 194]{grun_01}, $\approx\,1\times10^{-4}$~m$^{-2}$s$^{-1}$. The calculation is then similar to the case of ions. The spot size of the laser beam was several times larger than these particles, so the corresponding timescale is probably the lower limit.  

\section{Amorphization}
\label{appendix_amorphisation}
In Fig.~\ref{energy_evolution_fig} we can see that the spectral changes caused by various ions are similar. However, the trend representing H$^+$-irradiated olivine significantly deviates from the trends given by heavier ions at an energy density of about 100 J/cm$^2$. Our assumption is that the surface of the pellet exposed to the H$^+$ irradiation was subject to collapse of the crystalline structure (i.e. amorphization of the material). 

If we compare our set-up with that of \cite{brucato_04}, who studied crystal--amorphous changes of forsterite samples in IR spectroscopy due to H$^+$, He$^+$, carbon, and Ar$^{++}$ irradiation, we see that all our samples should be amorphized to a certain level. \cite{demyk_01} also did He$^+$ irradiation of crystalline olivine. They found that even at the lowest used fluence ($5\times10^{16}$~ions/cm$^2$, using 4 and 10~keV ions), layers of several tens of nanometres amorphized. It is thus probable that all our samples have some fraction of the volume amorphized, but only H$^+$-induced trends show a significant break. The reason for this may be that the different ions cause different amorphization levels. As \cite{demyk_04} pointed out, low-energy light ions and high-energy heavy ions have the greatest ability to amorphize silicates. Other combinations of energies and types of ions lead to the alteration of only a thin surface layer, whose contribution to the spectrum would  only be minor. An interplay between the penetration depths of different ions (see Table~\ref{srim_table}) and the thickness of the sample (size of the particles on the surface) can also influence the amorphization because, as noted by \cite{demyk_04}, the maximum damage happens at the penetration depth of the ions. If the grain is smaller than the penetration depth, it will show less damage than if the grain is larger. 

\begin{figure}
        \resizebox{\hsize}{!}{\includegraphics{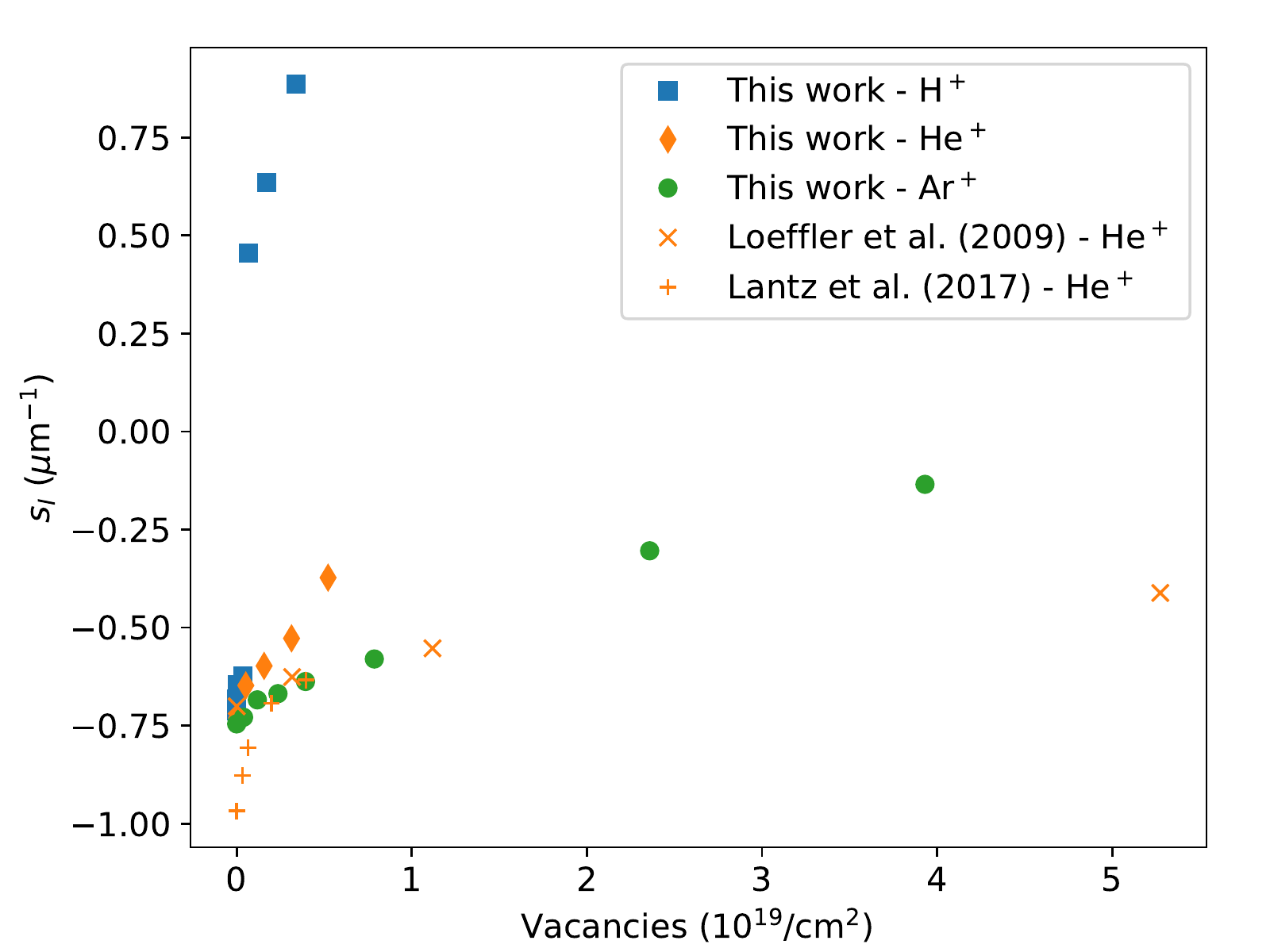}}
        \caption{Slope of 1~\textmu m band in olivines ($s_I$) as a function of the number of vacancies per cm$^2$ (taken from Table~\ref{srim_table}). The values of spectral slopes for the cited work are only estimations. H$^+$ stands for hydrogen, He$^+$ for helium, and Ar$^+$ for argon ions.}
        \label{vacancies_fig}
\end{figure}

\cite{brunetto_05} and \cite{brunetto_14} pointed out that the spectral changes relate to the nuclear elastic dose (i.e. the number of vacancies per cm$^2$). We thus plotted a similar graph to \citet[Fig.~8]{brunetto_14}. The dependence of the slope of the 1~\textmu m band, defined similarly to \cite{lazzarin_06} and calculated as a linear continuum from 540 to 920~nm of the spectra normalized at 550~nm, on the number of vacancies per cm$^2$ can be seen in Fig.~\ref{vacancies_fig}. Additionally, we estimated the slopes for the work of \cite{loeffler_09} and \cite{lantz_17}. Loeffler's spectra were measured from $\approx\,650$~nm. In that case the spectral slope is calculated from this shifted value and may thus be a bit offset. 

We can see that He$^+$ and Ar$^+$ irradiation show similar behaviour to that estimated from spectra in \cite{loeffler_09} and \cite{lantz_17}. It seems that the slope correlates with the number of vacancies and that for a high number of vacancies both types of irradiation reach an asymptotic value. On the other hand, our H$^+$ irradiation reaches very high and significantly different values of the spectral slope even at a low number of vacancies. Even this figure thus points to the distinct behaviour of the most  H$^+$-irradiated spectra. A complete understanding of the processes causing this discrepancy however requires transmission electron microscopy analyses and should be the subject of future work.

\section{Ratios of mid-infrared spectra}
\begin{figure}
        \resizebox{\hsize}{!}{\includegraphics{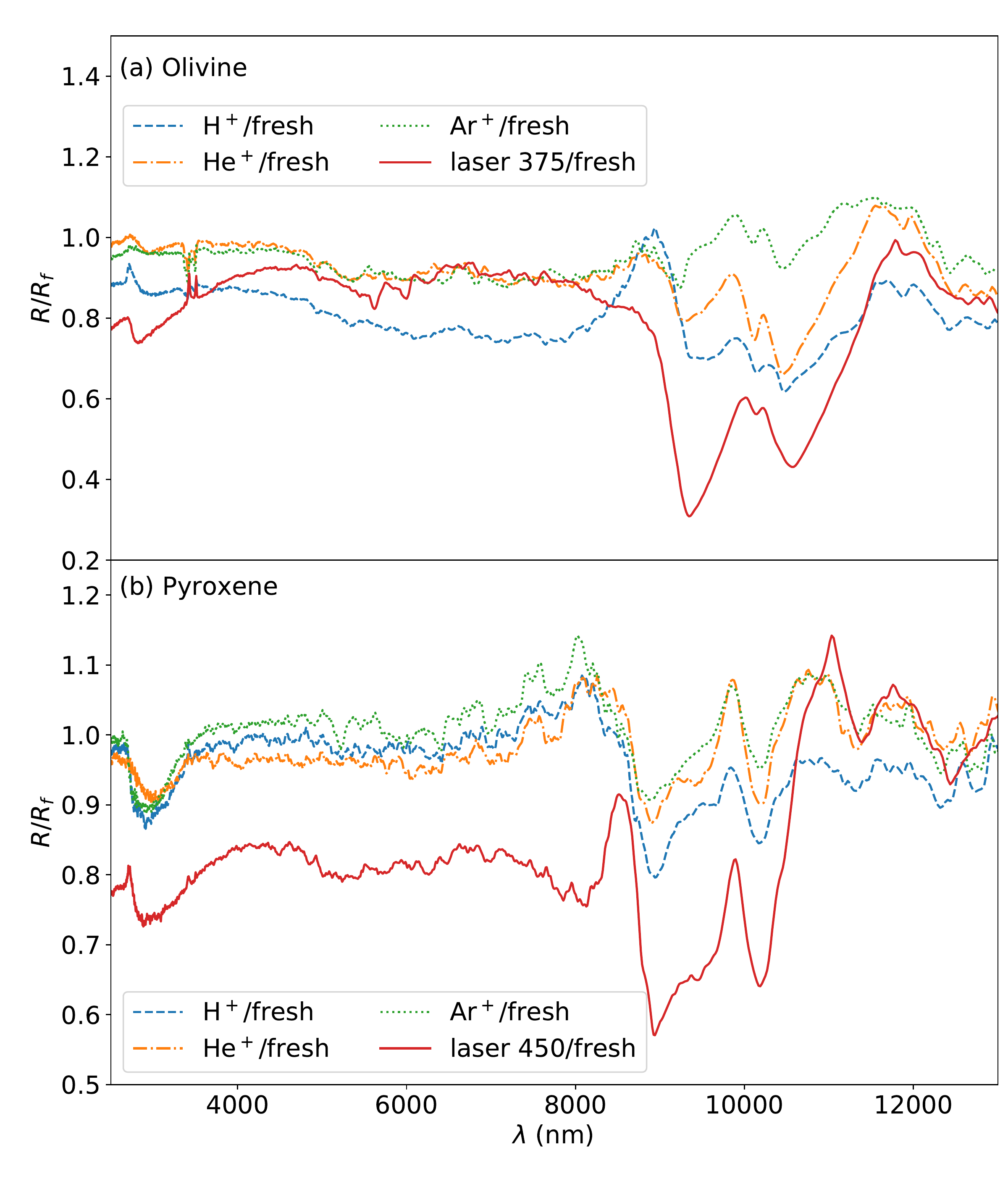}}
        \caption{Ratios of reflectance ($R$) of irradiated to fresh (f) spectra of olivine (OL) and pyroxene (OPX). $\lambda$ stands for wavelength. For ion irradiation,  the spectra with the highest irradiation were always used. For laser irradiation, we plotted the ratios using the 375 J/cm$^2$ (OL) and 450 J/cm$^2$ (OPX) cases, as they represent similar energy densities. H$^+$ stands for hydrogen, He$^+$ for helium, and Ar$^+$ for argon ions.}
        \label{appendix_mir_fig}
\end{figure}

Here we present ratios of irradiated and fresh MIR spectra affected by different irradiation types (see Fig.~\ref{appendix_mir_fig}). For both materials there is not a big change in the region up to 8~\textmu m. At longer wavelengths, changes in the positions, shapes, and intensities of the bands are obvious from the ratios. 

The reason for this behaviour may be the different scattering regime in these two regions of the spectra. Spectral features at shorter wavelengths than the Christiansen feature originate from the volume scattering, while features at the longer wavelengths, the reststrahlen bands, are caused by the surface scattering regime \citep{salisbury_91}. 

The difference at the shorter wavelengths of spectra of pyroxene irradiated by laser and by all other ions may thus be explained in the same manner as in the main text, that is by penetration depth (see Sect.~\ref{pen_depth_sect}). On the other hand, in the reststrahlen bands we obtain all the irradiation information from irradiated layers exclusively. (See the main text, Sects. \ref{mir_results} and \ref{mir_discussion}, for a deeper evaluation of the changes in the reststrahlen bands.)

\end{appendix}

\end{document}